\documentclass[10pt,journal,a4paper,twoside,twocolumn]{IEEEtran}
\IEEEoverridecommandlockouts
\usepackage{cite}
\usepackage{float}
\usepackage{stfloats}
\usepackage{fancyhdr}
\usepackage{color}
\usepackage[cmyk]{xcolor}
\usepackage{amsmath,amsfonts,amssymb}
\usepackage{graphicx}
\usepackage{algorithm}
\usepackage{algorithmic}
\usepackage{epsfig}
\usepackage{epstopdf}
\usepackage{bm}
\usepackage{multirow}
\usepackage{array}

\usepackage{bbm}
\usepackage[caption=false]{subfig}
\newtheorem{Lemma}{Lemma}
\newtheorem{Rem}{Remark}

\begin{document}
	\title{Reconfigurable Intelligent Sensing Surface enables Wireless Powered Communication Networks: Interference Suppression and Massive Wireless Energy Transfer
	}
	
	\author{\IEEEauthorblockN{
		Cheng Luo, \IEEEmembership{Graduate Student Member, IEEE}, Jie Hu, \IEEEmembership{Senior Member, IEEE}, Luping Xiang, \IEEEmembership{Member, IEEE} and Kun Yang, \IEEEmembership{Fellow, IEEE}
		}
			\\

		\thanks{Cheng Luo, Jie Hu, and Luping Xiang are with the School of Information and Communication Engineering, University of Electronic Science and Technology of China, Chengdu, 611731, China, email: chengluo@std.uestc.edu.cn; hujie@uestc.edu.cn; luping.xiang@uestc.edu.cn.}
		\thanks{Kun Yang is with the School of Computer Science and Electronic Engineering, University of Essex, Colchester CO4 3SQ, U.K., email: kunyang@essex.ac.uk.}
		\thanks{(Corresponding author: Jie Hu.)}
		
	}

	\maketitle
	
	\thispagestyle{fancy} 
	\lhead{} 
	\chead{} 
	\rhead{} 
	\lfoot{} 
	\cfoot{} 
	\rfoot{\thepage} 
	\renewcommand{\headrulewidth}{0pt} 
	\renewcommand{\footrulewidth}{0pt} 
	\pagestyle{fancy}

    \rfoot{\thepage} 

	\begin{abstract}
		Recently, a novel structures of reconfigurable intelligent surface (RIS) integrating both passive and active elements, termed reconfigurable intelligent sensing surface (RISS), efficiently addresses challenges in RIS channel estimation and mitigates issues related to multiplicative path loss by processing the signal at the RISS. In this paper, we propose a sensing-assisted wirelessly powered communication network (WPCN) that utilizes RISS's sensing capabilities to maximize the channel capacity in uplink wireless information transfer (WIT) and assist in massive wireless energy transmission (WET) for downlink. For the WIT in the uplink, the sensing information is utilized to design an interference suppression passive reflection phase shift for the RISS, and take the imperfect sensing results and sharp null into consideration, we also propose a robust scheme. For the WET in the downlink, the massive WET scheme is adopted and benefits from a period of sensing results. The massive WET scheme including beam selection and rotation order optimization to enhance the lower bound of energy harvest for massive users and optimize waiting costs. Numerical results demonstrate the optimal interference suppression threshold for uplink WIT and underscore the achieved fairness in downlink WET. Collectively, by utilizing sensing information, the uplink channel capacity is improved by 20\%, and the worst energy performance and waiting costs for massive WET are effectively optimized, with improvements ranging from 19\% to 59\% and 27\% to 29\%, respectively.
	\end{abstract}

\section{Introduction}

\IEEEPARstart{T}{he} future density of IoT devices is expected to escalate to tens or more per square meter \cite{intro_overviewonel1}, significantly impacting applications such as healthcare, environmental monitoring, smart homes, smart cities, autonomous vehicles, and national defense. This increased density, combined with the high-maintenance nature of these environments, renders frequent battery maintenance impractical. Consequently, it is imperative to exploit wireless technology to its fullest extent, not only for information transmission but also for energy delivery.

Wireless information and energy transfer is an emerging research area that utilizes radio waves for the joint purposes of wireless information transfer (WIT) and wireless energy transfer (WET). Numerous studies have highlighted the significance of WET \cite{WET_rectennas, WET_Modulation, WET2}, including the development of efficient rectennas \cite{WET_rectennas}, advancements in communications and signal design \cite{WET_Modulation}, and various practical transmitter/receiver architectures \cite{WET2}. However, in practical applications, an energy user typically requires substantially more power than an information user. Therefore, enhancing the efficiency of WET for energy users remains a critical challenge. 

Due to their ability to achieve high passive beamforming gain and compensate for significant radio frequency (RF) signal attenuation over long distances, reconfigurable intelligent surface (RIS) has found extensive applications in simultaneous wireless information and power transfer (SWIPT) \cite{RISSWIPT, RISSWIPT3, RISSWIPT4}, as well as in wirelessly powered communication networks (WPCNs) \cite{RISWPCN}. Moreover, their utility extends to various innovative domains, such as spatial electromagnetic wave frequency mixing \cite{frequencyadjustment2}, active RIS configurations \cite{activeIRS2} and integrated communication and sensing systems \cite{JointCommSensing}.

The integration of RIS into WIT and WET applications presents notable benefits, though challenges persist, particularly in obtaining channel state information (CSI) between RIS and associated primary bases/users \cite{CSIchallenge3}. The passive nature of RIS components complicates the acquisition of CSI, leading to increased pilot overhead proportional to the number of RIS elements and users, as noted in \cite{pilotoverhead1, pilotoverhead2}, thereby complicating traditional channel estimation methods.

Recent studies have addressed these challenges by developing innovative approaches for RIS channel estimation. Notable among there is the use of deep learning and compressed sensing techniques combined with randomly distributed active sensors to simplify the channel estimation process, significantly reducing pilot overhead \cite{csi_deeplearning}. Additionally, a novel approach involves placing a power sensor behind each RIS element to facilitate phase-coherent signal superposition at the receiver through the observation of interference \cite{sensingirs_dll}. Moreover, extensive empirical data aids in optimizing precoding and phase adjustments for RIS \cite{blindBF}, and location-based information systems enhance beamforming strategies \cite{locinformation}. A recent study also outlines a CSI-free approach for large-scale WET, directing energy efficiently across spatial domains using a beam rotation strategy\cite{Luo_MassiveWE}.

Moreover, there are a plethora of studies focusing on sensing-assisted communication, which introduce both new opportunities and challenges to sensing-assisted communication systems. In particular, \cite{SAKF} demonstrates the potential of AoA estimation-based Kalman filters in channel estimation for ISAC systems. The initial AoA estimated by the MUSIC method serves as prior information to refine the least squares (LS) CSI estimation, thereby enhancing the reliability of communication. A two-phase ISAC transmission protocol for the RIS-aided MIMO ISAC system was proposed in \cite{twophase}. This protocol utilizes limited CSI for WIT and rough sensing in the initial phase. Subsequently, the sensing information obtained in the first phase is utilized to improve both WIT and sensing in the second phase, enabling cooperative operation between WIT and sensing functions. Similarly, \cite{SensingAidedComm3} presents two deep learning networks designed for beam selection and power allocation, respectively. With the assistance of sensing results, these networks collectively maximize the communication channel capacity. Furthermore, \cite{SensingAidedComm4} introduces a sensing-assisted communication scheme leveraging RIS deployed on the surface of vehicles. This scheme aims to enhance both sensing and communication performance. The closed-form expression of the achievable rate under uncertain angle information is derived, significantly facilitating resource allocation in the considered systems.

The innovative concept of a reconfigurable intelligent sensing surface (RISS), which includes both active and passive elements, has also been introduced \cite{wqq_RISS, JSAC_RISS, RISS_luo}. This design significantly enhances localization capabilities through joint time-of-arrival (TOA) and direction-of-arrival (DOA) estimation, facilitating sophisticated sensing tasks without iterative algorithms \cite{wqq_RISS}. The RISS also plays a crucial role in dual-purpose uplink and downlink tasks within WPCNs, particularly in scenarios involving imperfect sensing \cite{JSAC_RISS, RISS_luo}. Partial related works are presented in Table \ref{table:relatedworks11}.

Despite these advances, research in RISS primarily centers on sensing applications. There is a compelling need to extend these findings to improve communication tasks supported by sensing data. Moreover, the potential of active elements within RISS to mitigate interference and enhance sensing-assisted communication remains largely untapped.
\begin{table*}[]
    \centering
	\caption{Relevant works on WIT/WET assisted by RIS and sensing.}
    \begin{tabular}{m{3 cm}<{\centering}|m{2.1 cm}<{\centering}|m{2.2 cm}<{\centering}|m{3.2 cm}<{\centering}|m{1.5 cm}<{\centering}|m{0.8 cm}<{\centering}}
    \hline
    \textbf{} &  \textbf{HAP control RIS}&  \textbf{Sensing scheme}& \textbf{Sensing-aided WIT/WET} & \textbf{Interference suppression} &\textbf{Robust design} \\
    \hline
    Our Proposed&No&DoA estimation, signal detection&WIT/WET&Yes&Yes\\
    \hline
	GPS-based\cite{locinformation}&Yes (small-scale)&Location&WIT&No&No\\
	\hline
	Two-phase scheme\cite{twophase}&Yes&Two-phase sensing & WIT&No&No\\
	\hline
	Sensing-aided Kalman filter (KF)\cite{SAKF}&Yes&DoA+KF-Based&WIT&No&No\\
	\hline
	Two-network scheme\cite{SensingAidedComm3}&-&Radar sensing&WIT(beam selection)&No&No\\
	\hline
	Vehicle sensing and communication\cite{SensingAidedComm4}&Yes&DoA estimation&WIT&No&Yes\\
	\hline
	RISS-based\cite{RISS_luo}&No&DoA estimation&WIT/WET&No&Yes\\
	\hline
    \end{tabular} \label{table:relatedworks11}
\end{table*}

This paper delineates our efforts to optimize the performance of WPCNs with RISS and sensing results support, dividing the challenge into uplink interference suppression and massive downlink WET. Our contributions are summarized as follows:
\begin{itemize}
	\item To optimize the performance of WPCNs, we decompose the challenge into two segments: an uplink WIT approach and a downlink WET strategy, both enhanced by sensing ability provided by RISS. Specifically, the uplink WIT benefits from instantaneous sensing, i.e., real-time DOA estimation and signal recognition, while the downlink WET benefits from a period of DOA estimations.
	\item In uplink WIT, signal processing at RISS contains suppresses interference and strengthens desired signals, thus optimizing SINR. The approach includes considerations for imperfect sensing, with empirical determination of optimal interference thresholds.
	\item In downlink WET, a beam rotation strategy for massive WET ensures equitable energy distribution. We determine the necessary beam count and optimize rotation sequences based on device directions obtained through a period of observations of uplink DOA estimation to minimize the waiting costs.
	\item Extensive experiments validate the efficiency of our proposed method, demonstrating robust performance even in complex Rician channels for WIT, with the uplink channel capacity improved by 20\%. For massive WET, the worst energy performance and waiting costs show improvements ranging from 19\% to 59\% and 27\% to 29\%, respectively.
\end{itemize}

The remaining sections of this paper are organized as follows: Section \ref{sec:systemmodel_pro} provides an overview of the system model, transmission protocol, and problem formulation. Section \ref{sec:WIT} presents the proposed interference suppression and robust design in the uplink WIT, while Section \ref{sec:WET} introduces the massive WET scheme in the downlink WET. Numerical results are detailed in Section \ref{sec:numericalresult}, and Section \ref{sec:conclusion} summarizes the findings and conclusions of this study.

\emph{Notation:} $\mathbf{I}_{M}$ represents the $M$-dimensional identity matrix, and $\mathbf{1}_{M}$ represents the $M\times 1$ matrix with all ones. The notation $[\cdot]_i$ and $[\cdot]_{i,j}$ refer to the $i$-th element of a vector and the $(i,j)$-th element of a matrix, respectively. The imaginary unit is denoted by $\mathbbm{i}=\sqrt{-1}$. The Euclidean norm and absolute value are denoted by $||\cdot||$ and $|\cdot|$, respectively. The function $\text{diag}\{\cdot\}$ creates a diagonal matrix. The mathematical expectation is denoted by $\mathbb{E}(\cdot)$. The operators $(\cdot)^{T}$, $(\cdot)^{\dagger}$ and $(\cdot)^{H}$ represent the transpose, conjugate and conjugate transpose, respectively. Finally, the notation $\mathcal{CN}$ represents the circularly symmetric complex Gaussian distribution.

\begin{figure*}
	\setlength{\abovecaptionskip}{0pt}
    \setlength{\belowcaptionskip}{0pt} 
	\centering
	\includegraphics[width=0.8\linewidth]{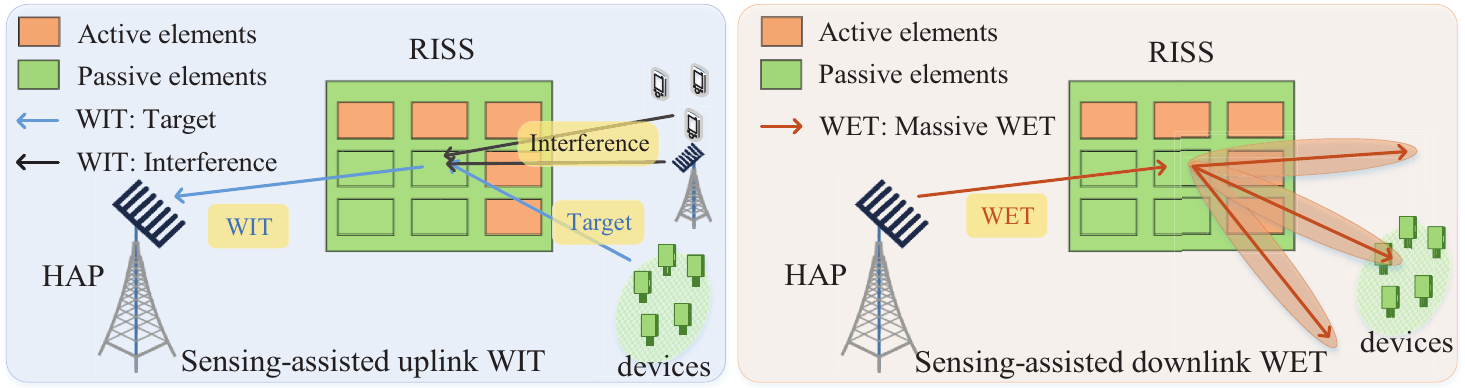}
	\caption{The system model of the proposed RISS-assisted WPCNs. The uplink WIT benefits from the real-time DOA estimation by the RISS, while the downlink WET benefits from a period of uplink transmission and DOA estimations.}
	\label{fig:systemmodel}
\end{figure*}

\section{System Model and Problem Formulation}\label{sec:systemmodel_pro}
This section presents the system architecture for the proposed RISS-assisted WPCNs. As illustrated in Fig. \ref{fig:systemmodel}, the system integrates a Hybrid Access Point (HAP) outfitted with a Uniform Linear Array (ULA) of $M$ antennas to serve multiple single-antenna Internet of Things (IoT) devices. The network enhancement is supported by an RISS that includes $(N_a + N)$ elements, where $N_a$ ($N_a \ll N$) is the number of active elements and $N = N_x \times N_y$ is the number of passive elements. The active elements, in general, are assumed to distribute in an L-array on one side of the RISS, as depicted in Fig. \ref{fig:systemmodel}. This structure remains the passive elements in the RISS as a uniform planar array (UPA), and may potentially mitigate coupling issues between passive and active elements\cite{RISS_luo}. The RISS fulfills dual roles of signal sensing and reflection. Note that the operational space contains not only the desired target signals but also interference from various sources such as other HAPs or non-cooperative users, as illustrates in Fig. \ref{fig:systemmodel}.

\subsection{Channel Model}
We assume quasi-static block fading for the line-of-sight (LoS) channels. The matrix $\mathbf{G}\in \mathbb{C}^{N\times M}$ represents the channel between the HAP and the RISS, whereas the vector $\mathbf{h}_{k}\in\mathbb{C}^{N\times 1}$ encapsulates the channel between the RISS and the $k^\text{th}$ IoT device or an interference source. Given the RISS's configuration as a UPA, the channels $\mathbf{G}$ and $\mathbf{h}_k$ are modeled as:
\begin{align}
	&\mathbf{G} = \boldsymbol{\alpha}(\vartheta_{G}, \varphi_{G})\boldsymbol{\beta}^T(\varpi_{G}), \label{eqn:channelG}\\
	&\mathbf{h}_{k} = \boldsymbol{\alpha}(\vartheta_{h,k}, \varphi_{h,k}),\label{eqn:channelh}
\end{align}
where the array response vectors $\boldsymbol{\alpha}(\vartheta, \varphi)$ and $\boldsymbol{\beta}(\varpi)$ can be expressed as  $\left[\boldsymbol{\alpha}(\vartheta, \varphi)\right]_n = e^{(\text{mod}(n,N_y)-1)\mathbbm{i}\vartheta}e^{(\lfloor n/N_x \rfloor - 1)\mathbbm{i}\varphi}$ and $\left[\boldsymbol{\beta}(\varpi)\right]_m = e^{(m-1)\mathbbm{i}\varpi}$, respectively. And the notation $[\cdot]_i$ and $[\cdot]_{i,j}$ refer to the $i$-th element of a vector and the $(i,j)$-th element of a matrix, respectively. The terms $\varphi$, $\vartheta$ and $\varpi$ are expressed as $\varphi = 2\pi d\cos(\phi^{\text{azi}})/\lambda = \pi\cos(\phi^{\text{azi}})$, $\vartheta = \pi\sin(\phi^{\text{azi}})\sin(\phi^{\text{ele}})$ and $\varpi=\pi\sin(\phi^{\text{dep}})$, respectively, with $d/\lambda = 1/2$ without loss of generality, where $d$ is the element spacing and $\lambda$ is the wavelength. $\phi^{\text{azi}}$, $\phi^{\text{ele}}$, and $\phi^{\text{dep}}$ represent the azimuth, elevation, and departure angles, respectively, which can be accurately estimated by the active elements of the RISS.

Furthermore, the accurate estimation of these incident angles is feasible owing to the active elements embedded within the RISS \cite{RISS_luo}. These active components are also capable of performing signal characterization tasks \cite{Modulationrecog}, such as distinguishing between interference and target signals. This dual functionality of angle estimation and signal identification enhances the RISS's capability to selectively manage the reflection of specific signals, thus optimizing the network performance\footnote{Although $N_a$ active elements handle specific signal processing tasks, the majority of the RISS structure remains passively reflective i.e., $N$ passive elements, maintaining a high passive gain while augmenting its overall intelligence.}.

\subsection{Signal Model}
In the uplink phase, the signal received at the HAP from the $k^\text{th}$ source is given by
\begin{align}
    y_k = \sqrt{\varrho_{R2H} \varrho_{U2R,k}}\mathbf{v}^T\mathbf{G}^T\boldsymbol{\Theta}\mathbf{h}_ks_k,\label{eqn:signalyk}
\end{align}
where $\varrho_{R2H}$ represents the path loss from the RISS to the HAP, and $\varrho_{U2R,k}$ denotes the path loss from the $k^\text{th}$ signal source to the RISS. Here, $\mathbf{v} \in \mathbb{C}^{M \times 1}$ is the receive beamforming vector applied at the HAP. The matrix $\boldsymbol{\Theta} = \text{diag}\{\beta_1, \dots, \beta_N\} \cdot \text{diag}\{\theta_1, \dots, \theta_N\}$ is the diagonal matrix representing the phase and amplitude settings of the RISS, with $\theta_i$ for all $i \in N$ and $\beta_i$ for all $i \in N$ as the phase shift and amplitude reflection coefficients, respectively. For simplicity and to ensure maximum reflection power, we set $\beta_i = 1$ for all $i \in N$. The symbol $s_k$ denotes the normalized signal from the $k^\text{th}$ source, such that $\mathbb{E}\{s_k^H s_k\} = 1$.

When multiple sources transmit signals simultaneously to the HAP, the aggregated received signal can be modeled as
\begin{align}
    y = \sum_{k=1}^K \sqrt{\varrho_{R2H} \varrho_{U2R,k}}\mathbf{v}^T\mathbf{G}^T\boldsymbol{\Theta}\mathbf{h}_ks_k + n_z,\label{eqn:totalreceived}
\end{align}
where $n_z \sim \mathcal{CN}(0, \sigma^2_0)$ is the additive white Gaussian noise, and $K$ denotes the number of active sources at any given moment. Consequently, the Signal-to-Interference-plus-Noise Ratio (SINR) for the $d^\text{th}$ IoT device is expressed as
\begin{align}
    \text{SINR}_d = \frac{\varrho_{R2H} \varrho_{U2R,d} \left|\mathbf{v}^T\mathbf{G}^T\boldsymbol{\Theta}\mathbf{h}_d\right|^2}{\sum_{k=1, k \neq d}^K \varrho_{R2H} \varrho_{U2R,k} \left|\mathbf{v}^T\mathbf{G}^T\boldsymbol{\Theta}\mathbf{h}_k\right|^2 + \sigma_0^2}.\label{eqn:sinr}
\end{align}

We focus on scenarios where only the signal from the $d^\text{th}$ device is considered as the target, while the other $\left(K-1\right)$ signals are treated as interference, without loss of generality. This setup is common in applications such as burst signal transmission, cell-free systems, and scenarios involving orthogonal or non-orthogonal pilot signals in massive IoT deployments\cite{cellfree, orth_nonorth}. Specifically, in a cell-free scenario, the device may experience interference from HAP and devices of other cells\cite{cellfree}. As for a single cell, the uncoordinated and random nature of the information traffic can be modeled as a Poisson process \cite{orth_nonorth}. Although orthogonal pilot reuse can effectively reduce the probability of collisions, they still persist. To this end, the scenario considered in this paper offers an effective solution to these issues, thereby simplifying the network configuration and improve performance of the communication system.

Similarly, the received energy of $k^{\text{th}}$ IoT device with energy beamforming $\mathbf{w}\in\mathbb{C}^{M\times1}$ in the downlink phase can be expressed as 
\begin{align}
	E = \varrho_{H2R} \varrho_{R2U,k}\left|\mathbf{h}_k^T\boldsymbol{\Theta}\mathbf{G}\mathbf{w}s_E\right|^2=\varrho_{H2U,k}\left|\mathbf{h}_k^T\boldsymbol{\Theta}\mathbf{G}\mathbf{w}\right|^2,
\end{align}
where $\varrho_{H2U,k} = \varrho_{H2R}\varrho_{R2U,k}$, $\varrho_{H2R}$ and $\varrho_{R2U,k}$ denote the path loss from HAP to the RISS and the RISS to the $k^\text{th}$ signal source, respectively. We assume that $s_E$ are independent white sequences from an arbitrary distribution with $\mathbb{E}\{|s_E|^2\}=1$ without loss of generality.
\subsection{Transmission Protocol}\label{sec:transmissionprotocols}

In this paper, we explore WPCNs for WET and WIT scenarios within periodically operating IoT devices. In this setup, WIT occurs from IoT devices to the HAP during the uplink phase, while WET occurs in the downlink to replenish energy expended by IoT devices for specific tasks such as environmental monitoring and data transmission. Note that sensing takes place at the beginning of WIT. Since the IoT device has a single antenna, the broadcast signals from the IoT device can be effectively detected by the active elements of the RISS. With the aid of both active and passive elements, sensing and reflection can be processed simultaneously. Benefiting from extensive literature on fast and single-shot DOA algorithms \cite{singlesnapshot1, singlesnapshot2}, the time required for sensing is considered negligible. Moreover, we propose deploying the RISS closer to the IoT devices to mitigate path loss between the RISS and the devices, thereby ensuring that the uplink signal received at the RISS maintains a sufficient SNR for accurate sensing and effective WET.

\begin{figure}
	\setlength{\abovecaptionskip}{0pt}
    \setlength{\belowcaptionskip}{0pt} 
	\centering
	\includegraphics[width=0.95\linewidth]{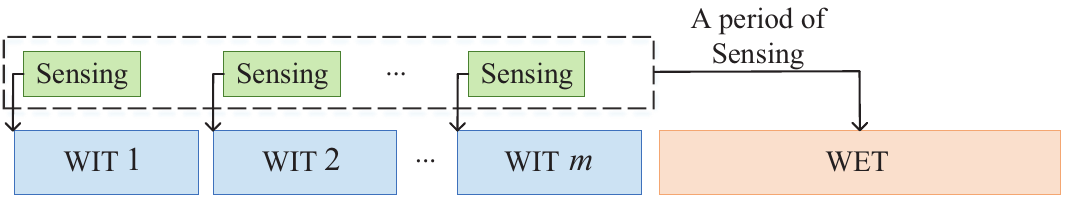}
	\caption{The frame structure of WIT and WET. We assume that there is a single target signal and multiple interference signals in each WIT frame.}
	\label{fig:framestruc}
\end{figure}

Fig. \ref{fig:framestruc} illustrates the sensing-aided frame structure for both WIT and WET. Specifically, WIT benefits from real-time sensing, i.e., DOA estimation and signal identification by the active elements in the RISS, which distinguishes between target signals and interference signals for interference suppression and target signal enhancement. Note that WET occurs after numerous WIT frames and utilizes the multiple DOA information obtained from a period of sensing\footnote{In particular, we can obtain the location of IoT devices from a period of sensing, which will be beneficial for our massive WET scheme.} for the massive WET beamforming scheme.

\section{Uplink Wireless Information Transfer}\label{sec:WIT}
The primary objective of this paper is to enhance the performance amidst a complex electromagnetic environment characterized by both target and interference sources within WPCNs. This objective necessitates a dual consideration: enhancing the SINR during WIT in the uplink and facilitating massive energy transfer during WET in the downlink. In this section, we first introduce the interference suppression design for WIT. Then, considering the possibility of imperfect sensing, we propose a robust design to address cases where DOA estimation is not accurate.

\subsection{The Orthogonality of angular}

To maximize the $\log_2\left(1+\text{SINR}_d\right)$ component in uplink WIT with the assistance of RISS, we first rewrite target signal power which corresponding to the numerator of Eq. \eqref{eqn:sinr} as
\begin{align}
	&\left|\mathbf{v}^T\mathbf{G}^T\boldsymbol{\Theta}\mathbf{h}_d\right|^2\nonumber\\
	\overset{(a)}{=}&\left|\mathbf{v}^T\boldsymbol{\beta}(\varpi_{G})\boldsymbol{\alpha}^T(\vartheta_{G}, \varphi_{G})\boldsymbol{\Theta}\boldsymbol{\alpha}(\vartheta_{h,d}, \varphi_{h,d})\right|^2\nonumber\\
	\overset{(b)}{=}&\left|\mathbf{v}^T\boldsymbol{\beta}(\varpi_{G})\text{diag}\{\boldsymbol{\Theta}\}^T\left(\boldsymbol{\alpha}(\vartheta_{G}, \varphi_{G})\circ\boldsymbol{\alpha}(\vartheta_{h,d}, \varphi_{h,d})\right)\right|^2,\label{eqn:numeratorofSINR}
\end{align}
where $(a)$ comes from Eq. \eqref{eqn:channelG} and Eq. \eqref{eqn:channelh}. $(b)$ is due to the fact $\mathbf{A}^T\boldsymbol{\Theta}\mathbf{B}=\text{diag}\{\boldsymbol{\Theta}\}^T(\mathbf{A}\circ\mathbf{B})$. To maximize Eq. \eqref{eqn:numeratorofSINR}, the receive beamforming and RISS phase shift can be derived as
\begin{align}
	&\mathbf{v}^T=\sqrt{P}\frac{\boldsymbol{\beta}^H(\varpi_G)}{||\boldsymbol{\beta}(\varpi_G)||},\label{eqn:alignv}\\
	&\text{diag}\{\boldsymbol{\Theta}\}=\left(\boldsymbol{\alpha}(\vartheta_{G}, \varphi_{G})\circ\boldsymbol{\alpha}(\vartheta_{h,d}, \varphi_{h,d})\right)^\dagger,\label{eqn:aligntheta}
\end{align}
where $P$ is the power of receive beamforming. And the maximum value of Eq. \eqref{eqn:numeratorofSINR} can be obtained as $PN^2M$ using Eq. \eqref{eqn:alignv} and Eq. \eqref{eqn:aligntheta}.

\begin{Lemma}\label{lemma:lemma1}
While Eq. \eqref{eqn:numeratorofSINR} reaches its maximum value (i.e., $PN^2M$), the power of the interference signal (corresponding to the component of $\left|\mathbf{v}^T\mathbf{G}^T\boldsymbol{\Theta}\mathbf{h}_k\right|^2, \forall k\in K, k\neq d$) can be expressed as
\begin{align}
	\left|\mathbf{v}^T\mathbf{G}^T\boldsymbol{\Theta}\mathbf{h}_k\right|^2=&\left|\mathbf{1}^T_N\boldsymbol{\alpha}_x(\varphi_{h,d}-\varphi_{h,k})\otimes\boldsymbol{\alpha}_y(\vartheta_{h,d}-\vartheta_{h,k})\right|^2, \label{eqn:interferencesignal2}
\end{align}
where $\boldsymbol{\alpha}_x(\varphi)=e^{(0:N_x-1)\mathbbm{i}\varphi}$, $\boldsymbol{\alpha}_y(\vartheta)=e^{(0:N_y-1)\mathbbm{i}\vartheta}$, and $\boldsymbol{\alpha}_x(\varphi)\otimes\boldsymbol{\alpha}_y(\vartheta)=\boldsymbol{\alpha}(\vartheta, \varphi)$. Thus, the requirements for interference elimination (i.e., $\left|\mathbf{v}^T\mathbf{G}^T\boldsymbol{\Theta}\mathbf{h}_k\right|^2=0$) are
\begin{align}
	&\left|\cos(\phi^{\text{azi}}_{h,d})-\cos(\phi^{\text{azi}}_{h,k})\right|=\frac{2n}{N_x},n\in \mathbb{N}^+,\label{eqn:orthangle1}\\
	&\left|\sin(\phi^{\text{azi}}_{h,d})\sin(\phi^{\text{ele}}_{h,d})-\sin(\phi^{\text{azi}}_{h,k})\sin(\phi^{\text{ele}}_{h,k})\right|=\frac{2n}{N_y},n\in \mathbb{N}^+,\label{eqn:orthangle2}
\end{align}
\end{Lemma}
where $\mathbb{N}^+$ denotes the positive integer set. We refer to the angles that satisfy the aforementioned equations as orthogonal angles.

\begin{IEEEproof}
	Please refer to Appendix \ref{app:A} for detailed proof.
\end{IEEEproof}

\begin{figure}
	\setlength{\abovecaptionskip}{0pt}
    \setlength{\belowcaptionskip}{0pt} 
	\centering
	\includegraphics[width=0.85\linewidth]{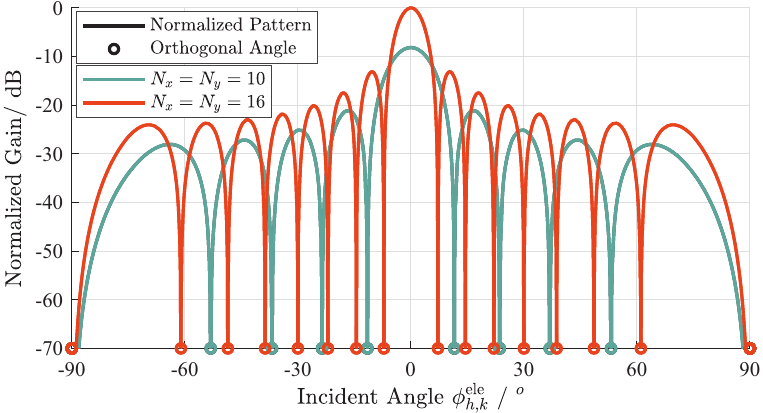}
	\caption{Normalized gain and orthogonal angles of RISS reflection patterns with $N=100$ and $N=256$ passive elements. The values are fixed as $\phi^{\text{azi}}_{h,d} = \phi^{\text{azi}}_{h,k} = \pi/2$ and $\phi^{\text{ele}}_{h,d} = 0$, illustrating the suppression of interference at orthogonal angles.}
	\label{fig:orthangle}
\end{figure}

\begin{Rem}\label{remark:1}
	Lemma \ref{lemma:lemma1} demonstrates that interference signals from specific positions relative to RISS (in compliance with Eq. \eqref{eqn:orthangle1} and \eqref{eqn:orthangle2}) are automatically suppressed while maximizing the target signal. This phenomenon is further elucidated in Fig. \ref{fig:orthangle}. By setting $\phi^{\text{azi}}_{h,d}=\phi^{\text{azi}}_{h,k}=\pi/2$ and $\phi^{\text{ele}}_{h,d} = 0$, we demonstrate a scenario where both the target signal and interference signal reside on the same plane\footnote{Note that the legend in Fig. \ref{fig:orthangle} is represented as a combination of two elements: shape and color, in order to simplify its presentation.}. Interference signals originating from orthogonal angles are efficiently suppressed, thereby significantly enhancing the SINR at the HAP for the target signal. Moreover, the number of orthogonal angles increases with the augmentation of passive elements in RISS (e.g., in the scenario depicted in Fig. \ref{fig:orthangle}, the number of orthogonal angles equals $N_y$).
\end{Rem}

Due to the inherent mobility and randomness of both target and interference sources, achieving perfect angle orthogonality is challenging. Future sections will explore strategies to mitigate interference from various angles effectively.

\subsection{The General Solution and Robust Design}\label{sec:robustdesign}
In this section, we consider signals from random angle incidence, viewing the RISS as a spatial filter that selectively permits specific signals and eliminate the interference signals based on estimated angle information from the active elements, thereby maximizing the SINR. With this objective in mind, the target signal in Eq. \eqref{eqn:numeratorofSINR} can be rewritten as
\begin{align}
	&\left|\mathbf{v}^T\boldsymbol{\beta}(\varpi_{G})\boldsymbol{\alpha}^T(\vartheta_{G}, \varphi_{G})\boldsymbol{\Theta}\boldsymbol{\alpha}(\vartheta_{h,d}, \varphi_{h,d})\right|^2\nonumber\\
	=&\left|\mathbf{c}^T\boldsymbol{\alpha}(\vartheta_{h,d},\varphi_{h,d})\right|^2\nonumber\\
	=&\mathbf{c}^T\boldsymbol{\alpha}(\vartheta_{h,d},\varphi_{h,d})\boldsymbol{\alpha}(\vartheta_{h,d},\varphi_{h,d})^H\mathbf{c}^\dagger\nonumber\\
	=&\mathbf{c}^T\boldsymbol{\mathcal{A}}_d\mathbf{c}^\dagger\nonumber\\
	=&\text{trace}(\boldsymbol{\mathcal{C}}\boldsymbol{\mathcal{A}}_d),\label{eqn:calA}
\end{align}
where $\mathbf{c}^T=\mathbf{v}^T\boldsymbol{\beta}(\varpi_{G})\boldsymbol{\alpha}^T(\vartheta_G,\varphi_G)\boldsymbol{\Theta}$, $\boldsymbol{\mathcal{C}}=\mathbf{c}^\dagger\mathbf{c}^T$, $\boldsymbol{\mathcal{A}}_d=\boldsymbol{\alpha}(\vartheta_{h,d},\varphi_{h,d})\boldsymbol{\alpha}(\vartheta_{h,d},\varphi_{h,d})^H$. Similarly, we can rewrite Eq. \eqref{eqn:interferencesignal2} as
\begin{align}
	\left|\mathbf{v}^T\mathbf{G}^T\boldsymbol{\Theta}\mathbf{h}_k\right|^2=\text{trace}(\boldsymbol{\mathcal{C}}\boldsymbol{\mathcal{A}}_k), \forall k\in K, k\neq d.\label{eqn:calB}
\end{align}

Eq. \eqref{eqn:calA} and \eqref{eqn:calB} provide the expressions for the target signal and interference signal, respectively. After obtaining the angle information of the incident signals using the $N_a$ active sensing elements of the RISS, our objective is to design the reflection phase shifts of the passive elements of the RISS based on the sensing information\footnote{Note that while the integration of active elements results in increased energy consumption, it also empowers the RISS to operate autonomously, thereby obviating the need for the feedback link typically required between the RIS and HAP in traditional RIS-based systems. Furthermore, during the downlink WET phase, the HAP can supply energy to the RISS, thereby preparing it for the subsequent sensing tasks in the upcoming WIT phase. However, due to space limitations, the solution for HAP charging the RISS is deferred to future work.}. This is done in such a way that the RISS can suppress the interference signals while simultaneously enhancing the target signal. Therefore, to maximize the target signal and eliminate the interference signals, the objective can be formulated as

\begin{align}
		\text{(P1):  }\max_{\mathbf{v},\boldsymbol{\Theta}} \,\,&\text{trace}(\boldsymbol{\mathcal{C}}\boldsymbol{\mathcal{A}}_d)\label{eqn:eqnP1}\\
		\text{s.t.}\,\,\,\,&\text{diag}(\boldsymbol{\mathcal{C}})=\left|\mathbf{v}^T\boldsymbol{\beta}(\varpi_{G})\right|^2\mathbf{1}_N,\tag{\ref{eqn:eqnP1}a}\label{eqn:eqnP1consA}\\
		&\text{trace}(\boldsymbol{\mathcal{C}})\leq MNP,\tag{\ref{eqn:eqnP1}b}\label{eqn:eqnP1consB}\\
		&\text{trace}(\boldsymbol{\mathcal{C}}\boldsymbol{\mathcal{A}}_k)=0,\forall k\in K, k\neq d,\tag{\ref{eqn:eqnP1}c}\label{eqn:eqnP1consC}\\
		&\boldsymbol{\mathcal{C}} = \mathbf{c}^\dagger\mathbf{c}^T,\tag{\ref{eqn:eqnP1}d}\label{eqn:eqnP1consD}\\
		&\mathbf{c}^T=\mathbf{v}^T\boldsymbol{\beta}(\varpi_{G})\boldsymbol{\alpha}^T(\vartheta_G,\varphi_G)\boldsymbol{\Theta}\tag{\ref{eqn:eqnP1}e}\label{eqn:eqnP1consE},\\
		&\text{Rank}(\boldsymbol{\mathcal{C}}) = 1,\tag{\ref{eqn:eqnP1}f}\label{eqn:eqnP1consF}
\end{align}
where the objective $\text{(P1)}$ is the power of target signals, and Eq. \eqref{eqn:eqnP1consA}-\eqref{eqn:eqnP1consC} come from RISS passive phase shift constrain, transmit power constrain and interference elimination.

Obviously, Eq. \eqref{eqn:alignv} is still the optimal beamforming at HAP to maximize the component $\mathbf{v}^T\boldsymbol{\beta}(\varpi_{G})$. We divide $\boldsymbol{\Theta}$ into $\boldsymbol{\Theta}=\boldsymbol{\Theta}_G\boldsymbol{\Theta}_h$ for clarity, where $\boldsymbol{\Theta}_G$ and $\boldsymbol{\Theta}_h$ are both diagonal matrix. Since the rank-one constraint is non-convex, we apply semidefinite relaxation (SDR) to relax this constraint, result in
\begin{align}
	\text{(P1.1):  }\max_{\boldsymbol{\Theta}_h} \,\,\,&\text{trace}(\boldsymbol{\mathcal{C}}\boldsymbol{\mathcal{A}}_d)\label{eqn:eqnP1p1}\\
	\text{s.t.}\,\,\,\,\,&\text{diag}(\boldsymbol{\mathcal{C}})=MP\mathbf{1}_N,\tag{\ref{eqn:eqnP1p1}a}\label{eqn:eqnP1p1consA}\\
	&\mathbf{c}^T=\sqrt{MP}\mathbf{1}_N^T\boldsymbol{\Theta}_h,\tag{\ref{eqn:eqnP1p1}b}\label{eqn:eqnP1p1consB}\\
	&\text{trace}(\boldsymbol{\mathcal{C}}\boldsymbol{\mathcal{A}}_k)\leq \tau_k,\forall k\in K,k\neq d,\tag{\ref{eqn:eqnP1p1}c}\label{eqn:eqnP1p1consC}\\
	&\boldsymbol{\mathcal{C}}\succeq 0,\tag{\ref{eqn:eqnP1p1}d}\label{eqn:eqnP1p1consD}\\
	&\eqref{eqn:eqnP1consB}, \eqref{eqn:eqnP1consD}, \tag{\ref{eqn:eqnP1p1}e}\label{eqn:eqnP1p1consE}
\end{align}
where \eqref{eqn:eqnP1p1consA} comes from Eq. \eqref{eqn:alignv}, \eqref{eqn:eqnP1p1consB} comes from setting $\boldsymbol{\Theta}_G=\boldsymbol{\alpha}^\dagger(\vartheta_G,\varphi_G)$ and $\boldsymbol{\alpha}^T(\vartheta_G,\varphi_G)\boldsymbol{\Theta_G}=\mathbf{1}_N^T$. We introduced a reasonable threshold, i.e, $\tau_k,\forall k\in K, k\neq d$, within constraint \eqref{eqn:eqnP1p1consC} to assess the extent of interference suppression, corresponding to constraint \eqref{eqn:eqnP1consC}. Moreover, owing to the elimination of the rank-one constraint of $\boldsymbol{\mathcal{C}}$, additional steps i.e., iterative rank minimization algorithm (IRM) or Gaussian randomization are needed to construct a rank-one solution from a higher-rank solution.

Specifically, upon obtaining the preliminary solution from $\text{(P1.1)}$, denoted as $\boldsymbol{\mathcal{C}}_0$, it is essential to determine the rank of $\boldsymbol{\mathcal{C}}_0$. A rank-one solution in $\boldsymbol{\mathcal{C}}_0$ indicates the optimal solution has been achieved. In contrast, the IRM algorithm is employed for rank-one recovery. Note that IRM is an iterative algorithm, we illustrate the process by considering one iteration of the algorithm as an example. We commence by executing eigenvalue decomposition on $\boldsymbol{\mathcal{C}}_0$, denoting its eigenvalues in ascending order as $[\lambda_1,\cdots,\lambda_N]$. Hence, the eigenvalues can be expressed as
\begin{align}
	\text{diag}\left(\lambda_1,\cdots,\lambda_{N-1}\right)=\mathbf{V}_0^H\boldsymbol{\mathcal{C}}_0\mathbf{V}_0,
\end{align}
where $\mathbf{V}$ denotes the matrix composed of eigenvectors corresponding to $[\lambda_1,\cdots,\lambda_{N-1}]$. Thus, the rank one constrain can be converted to
\begin{align}
	r\mathbf{I}_{N-1}-\mathbf{V}_0^H\boldsymbol{\mathcal{C}}_0\mathbf{V}_0 \succeq \mathbf{0},
\end{align}
where $r$ is the positive relaxation variable. As $r$ approaches zero, the undesired eigenvalue $\text{diag}(\lambda_1,\cdots,\lambda_{N-1})$ are forced to become zeros, which implies that $\boldsymbol{\mathcal{C}}_0$ satisfies the rank-one constraint, and the iteration is terminated. Thus, with the assistance of IRM algorithm, the objective $\text{(P1.1)}$ is converted into

\begin{align}
	\text{(P1.2):  }\min_{\boldsymbol{\Theta}_{h,l}, r} \,\,\,&-\text{trace}(\boldsymbol{\mathcal{C}}_l\boldsymbol{\mathcal{A}}_d)+\epsilon_l r\label{eqn:eqnP1p2}\\
	\text{s.t.}\,\,\,\,\,&\text{diag}(\boldsymbol{\mathcal{C}}_l)=MP\mathbf{1}_N,\tag{\ref{eqn:eqnP1p2}a}\label{eqn:eqnP1p2consA}\\
	&\mathbf{c}^T_l=\sqrt{MP}\mathbf{1}_N^T\boldsymbol{\Theta}_{h,l},\tag{\ref{eqn:eqnP1p2}b}\label{eqn:eqnP1p2consB}\\
	&\text{trace}(\boldsymbol{\mathcal{C}}_l\boldsymbol{\mathcal{A}}_k)\leq \tau_k,\forall k\in K,k\neq d,\tag{\ref{eqn:eqnP1p2}c}\label{eqn:eqnP1p2consC}\\
    &r\mathbf{I}_{N-1}-\mathbf{V}_l^H\boldsymbol{\mathcal{C}}_l\mathbf{V}_l \succeq \mathbf{0},\tag{\ref{eqn:eqnP1p2}d}\label{eqn:eqnP1p2consD}\\
	&\boldsymbol{\mathcal{C}}_l\succeq 0,\tag{\ref{eqn:eqnP1p2}e}\label{eqn:eqnP1p2consE}\\
	&\eqref{eqn:eqnP1consB}, \eqref{eqn:eqnP1consD}, \tag{\ref{eqn:eqnP1p2}f}\label{eqn:eqnP1p2consF}
\end{align}
where $\boldsymbol{\mathcal{C}}_l$ denotes the solution in the $l^\text{th}$ iteration, and $\epsilon_l$ denotes a gradually increasing auxiliary variable. The IRM algorithm based solution is summarized in Algorithm \ref{alg:1}. In each iteration of the IRM algorithm, the problem $\text{(P1.2)}$ is solved until either the maximum number of iterations is reached or the positive relaxation variable $r$ approaches zero. Since $\text{(P1.2)}$ is a convex optimization problem, standard optimization methods, such as interior-point methods, can be employed to solve it during each iteration.

\begin{algorithm}[t]
	\small
	\caption{IRM algorithm based solution for $\text{(P1.2)}$.}
	\begin{algorithmic}[1]\label{alg:1}
		\REQUIRE~\ $\boldsymbol{\mathcal{A}}_d$, $M$, $P$, $\tau_k,\forall k\in K, k\neq d$.
		\ENSURE~\ Optimal solution $\boldsymbol{\Theta}_{h}$.
		\STATE Initialize $\epsilon_0=4$, $l=0$, $\varsigma=1.5$
		\STATE Obtain the preliminary solution $\boldsymbol{\mathcal{C}}_0$ from $\text{(P1.1)}$;
		\REPEAT
		\STATE Obtain the matrix $\mathbf{V}_l$ from eigenvalue decomposition;
		\STATE Solve the problem $\text{(P1.2)}$ to obtain $\boldsymbol{\mathcal{C}}_{l+1}$ and $r$;
		\STATE Update $\epsilon_{l+1} = \epsilon_{l}^\varsigma$;
		\STATE $l=l+1$;
		\UNTIL{$r$ reaches a sufficiently small value or reaches the maximum iterations.}
		\STATE Obtain the optimal solution $\boldsymbol{\Theta}_{h}$ from eigenvalue decomposition of $\boldsymbol{\mathcal{C}}_l$.
	\end{algorithmic}
\end{algorithm}

For clarity, let $\boldsymbol{\Theta}_\text{ALI}$ denote the RISS phase shift specifically engineered to maximize target signals, as defined in Eq. \eqref{eqn:aligntheta}. Conversely, $\boldsymbol{\Theta}_\text{ELI}$ represents the phase shift designed to eliminate interference, described by
\begin{align}
    \boldsymbol{\Theta}_\text{ELI} = \boldsymbol{\Theta}_G\boldsymbol{\Theta}_h = \text{diag}\left(\boldsymbol{\alpha}^\dagger(\vartheta_G,\varphi_G)\right)\boldsymbol{\Theta}_h,
\end{align}
where $\boldsymbol{\Theta}_h$ aligns with the principal eigenvector of the matrix $\boldsymbol{\mathcal{C}}$\cite{RISSWIPT3}, which is obtained from $\text{(P1.2)}$.

\begin{figure}
	\setlength{\abovecaptionskip}{0pt}
    \setlength{\belowcaptionskip}{0pt} 
    \centering
    \includegraphics[width=0.85\linewidth]{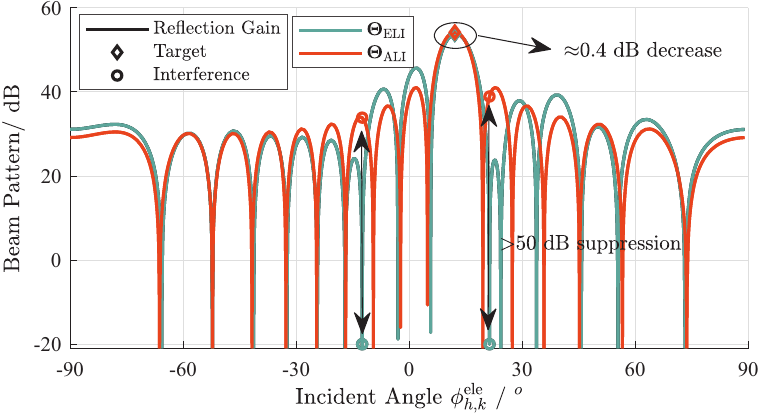}
    \caption{Comparison of $\boldsymbol{\Theta}_\text{ALI}$ and $\boldsymbol{\Theta}_\text{ELI}$ schemes showing reflection gains with $\tau_k=-20$ dB, $\forall k \in K, k \neq d$, for $N = 256$ and $M=4$. The target signal incidents at $12.1^\circ$, with interference signals at $21.3^\circ$ and $-12.5^\circ$.}
    \label{fig:eliandalign}
\end{figure}

\begin{Rem}\label{remark:2}
Objective $\text{(P1.2)}$ aims to derive a RISS phase shift, $\boldsymbol{\Theta}_\text{ELI}$, that suppresses interference signals beyond a predefined threshold $\tau_k, \forall k \in K, k \neq d$, with minimal impact on the reflection gain of target signals. Fig. \ref{fig:eliandalign} illustrates that while $\boldsymbol{\Theta}_\text{ELI}$ effectively mitigates interference, $\boldsymbol{\Theta}_\text{ALI}$ may retain interference within its sidelobes, potentially reducing SINR. Moreover, the mainlobe maintains a specific width, whereas the interference null is notably sharp. Accurate angle estimation allows the null to distinctively resolve between target and interference signals, enhancing signal separation. Conversely, errors in angle estimation may compromise interference suppression efficacy, though target signal enhancement remains relatively robust to such inaccuracies.
\end{Rem}

\begin{algorithm}[t]
	\small
    \caption{Threshold Search Algorithm in the Interval $[\Delta_s$, $\Delta_e]$.}
    \begin{algorithmic}[1]\label{alg:2}
        \REQUIRE Initial scan interval [$\Delta_s$, $\Delta_e$], scan lengths $l_c$ and $l_f$.
        \ENSURE Optimal threshold $\gamma_{\text{peak},i},\forall i\in N_p$ for beam rotation.

        \textcolor{gray}{// Rough Phase}
        \STATE Compute $\delta_c = \frac{\Delta_e - \Delta_s}{l_c}$;
        \STATE Set $\delta = \delta_c$;
        \FOR {$\gamma$ in $[\Delta_s:\delta:\Delta_e]$}
            \STATE Compute $B_{\text{obs}} = \pi/2 - B_{\gamma, L}(\pi/2)$ or $\pi/2$;
            \STATE Set $\theta_s = B_{\text{obs}}$, $\theta_e = -\delta_0$;
            \REPEAT
                \STATE Identify $\theta \in [\theta_s, \theta_e]$ minimizing $\left|F_{B_\text{obs}}(\theta)\right|^2 - \gamma$;
                \STATE Update $B_\text{obs} = \theta_s = \theta - B_{\gamma, L}(\theta)$;
            \UNTIL{$B_\text{obs} \leq B_{\gamma, R}(0)$ or $B_\text{obs} \leq 0$};
            \STATE Append $d_{\text{res}} = [d_{\text{res}}, \min\left\{\left|B_{\text{obs}}\right|, \left|B_{\text{obs}} - B_{\gamma, R}(0)\right|\right\}]$;
        \ENDFOR
        \STATE Determine $\gamma_{\text{peak},i},\forall i\in N_p$ from peaks of $1/d_\text{res}$;
        \STATE Refine range $[\Delta_s, \Delta_e]$ into sub-ranges $[\Delta_{s, i}, \Delta_{e,i}],\forall i\in N_p$, where $\Delta_{s, i} = \gamma_{\text{peak},i} - \delta$ and $\Delta_{e, i} = \gamma_{\text{peak},i} + \delta$.

        \textcolor{gray}{// Fine Phase}
        \FOR{each range $[\Delta_{s, i}, \Delta_{e,i}],\forall i\in N_p$}
            \REPEAT
                \STATE Compute $\delta = \frac{\Delta_{e,i} - \Delta_{s,i}}{l_f}$;
                \STATE Set $\delta = \delta_f$;
                \FOR{$\gamma$ in $[\Delta_{s, i}:\delta: \Delta_{e,i}]$}
                    \STATE \textit{Repeat steps 4-10.}
                \ENDFOR
                \STATE Determine $\gamma_{\text{peak},i},\forall i\in N_p$ from peaks of $1/d_\text{res}$;
                \STATE Adjust $\Delta_{s, i} = \gamma_{\text{peak},i} - \delta$, $\Delta_{e, i} = \gamma_{\text{peak},i} + \delta$;
            \UNTIL{convergence or maximum iterations are reached};
            \STATE Finalize optimal $\gamma_{\text{peak},i},\forall i\in N_p$.
        \ENDFOR
    \end{algorithmic}
\end{algorithm}

Building upon Remark \ref{remark:2}, in the presence of sensing errors, the interference suppression strategy utilizing imperfect sensing information may become less effective, as the nulls formed by the RISS reflection phase shift matrix are typically very narrow. To this end, we extends to explore interference suppression under angle estimation errors. A pragmatic approach involves broadening the nulls, imposing constraints over a range $[\hat{\phi}^\text{ele}_{h,k}-\delta_{h,k}, \hat{\phi}^\text{ele}_{h,k}+\delta_{h,k}], k \in K, k \neq d$, where $\hat{\phi}^\text{ele}_{h,k}$ represents the estimated angle for the $k^\text{th}$ interference signal. Here, it is assumed that $\phi^{\text{azi}}_{h,d} = \phi^{\text{azi}}_{h,k} = \pi/2$, positioning all signals within the same plane and disregarding estimation errors for $\phi^{\text{azi}}_{G}$, $\phi^{\text{ele}}_{G}$, and $\phi^{\text{dep}}_{G}$ due to the static nature of the HAP and RISS.

The interference range is segmented into an $L$ grid by the step ${2\delta_{h,k}}/{(L-1)}$, denoted by $\phi_{k,l}, k \in K, k \neq d, l \in L$. The reformulated objective $\text{(P2)}$ is given by
\begin{align}
    \text{(P2):  }\max_{\boldsymbol{\Theta}_h} \,\,\,&\text{trace}(\boldsymbol{\mathcal{C}}\boldsymbol{\mathcal{A}}_d)\label{eqn:eqnP2}\\
    \text{s.t.}\,\,\,\,\,&\text{diag}(\boldsymbol{\mathcal{C}})=MP\mathbf{1}_N,\tag{\ref{eqn:eqnP2}a}\label{eqn:eqnP2consA}\\
    &\mathbf{c}^T=\sqrt{MP}\mathbf{1}_N^T\boldsymbol{\Theta}_h,\tag{\ref{eqn:eqnP2}b}\label{eqn:eqnP2consB}\\
    &\text{trace}(\boldsymbol{\mathcal{C}}\boldsymbol{\mathcal{A}}_{k,l})\leq \tau_{k}, k \in K,k \neq d, l \in L,\tag{\ref{eqn:eqnP2}c}\label{eqn:eqnP2consC}\\
	&\boldsymbol{\mathcal{C}}\succeq 0,\tag{\ref{eqn:eqnP2}d}\label{eqn:eqnP2consD}\\
    &\eqref{eqn:eqnP1consB}, \eqref{eqn:eqnP1consD}, \tag{\ref{eqn:eqnP2}e}\label{eqn:eqnP2consE}
\end{align}
where $\boldsymbol{\mathcal{A}}_{k,l}, k \in K, k \neq d, l \in L$ is calculated based on $\phi_{k,l}$. The robust design efforts are focused on interference signals, leveraging inherent robustness within the mainlobe of the target signal. The solution to $\text{(P2)}$, denoted as $\boldsymbol{\Theta}^R_\text{ELI}$, is achieved using the IRM algorithm and also can be solved by standard optimization method, such as the interior-point method during each iteration.

Note that the computational complexity of the proposed algorithm can be divided into two main components: the sensing complexity and the complexity associated with solving the optimization problems $\text{(P1.2)}$ and $\text{(P2)}$. The sensing complexity arises from the subspace-based super-resolution algorithm employed for DOA estimation of the active elements, which incurs a computational complexity of $\mathcal{O}\left(T_{s}N_a^3\right)$, where $T_s$ is the number of snapshots. In contrast, the complexity of solving each optimization problem $\text{(P1.2)}$ and $\text{(P2)}$ using the interior-point method is $\mathcal{O}\left(N^{3.5} \log \frac{1}{\zeta}\right)$, where $\zeta$ denotes the prescribed accuracy \cite{luo2010semidefinite}. Let $I_{IRM}$ represent the number of iterations required for the IRM algorithm, each of which needs eigenvalue decomposition and introduces $\mathcal{O}(N^3)$ computational complexity. Consequently, the total computational complexity of the proposed algorithm can be expressed as $\mathcal{O}\left(I_{IRM} \left(N^{3.5} \log \frac{1}{\zeta}+N^3\right) + T_sN_a^3\right)$. Given that $N \gg N_a$, the computational complexity can be further simplified to $\mathcal{O}\left(I_{IRM} N^{3.5} \log \frac{1}{\zeta}\right)$.

\section{Downlink Massive Energy Transfer}\label{sec:WET}
Benefits from a period of sensing as illustrates in Fig. \ref{fig:framestruc}, we obtain the locations of IoT devices. In this section, we aim to design a massive WET scheme, including beam selection and optimal rotation order to enhance the lower bound of energy harvest and minimize the waiting cost during the WET.

\subsection{The Beam Rotation Scheme for Massive Wireless Energy Transfer}
Previous studies \cite{RISS_luo, Luo_MassiveWE} highlight that traditional beamforming techniques for RIS such as \cite{wqq_WET} do not uniformly distribute energy among multiple IoT devices, and perfect CSI for RIS is often unavailable. In response, we introduce a novel beamforming scheme that divides the service area into multiple beam coverage intervals, ensuring complete coverage within a single period through beam rotation strategy.

When optimizing downlink WET in a specific direction $\hat{\vartheta}_{h}$, including adopting the transmit energy beamforming and RISS phase shift as Eq. \eqref{eqn:alignv} and Eq. \eqref{eqn:aligntheta}. Thus, the IoT device located at ${\vartheta}_{h,d} = \hat{\vartheta}_{h}$ can receive the maximum energy gain, as depicted in Lemma \ref{lemma:lemma1}. For the energy performance of other IoT devices, we first define the normalized beam pattern relationship with the direction as per \cite{Beamwidth}
\begin{align}
    F_{\hat{\vartheta}_{h}}(\omega) = \frac{\sin\left(\frac{N\pi d}{\lambda}(\sin(\omega) - \sin(\hat{\vartheta}_{h}))\right)}{N\sin\left(\frac{\pi d}{\lambda}(\sin(\omega) - \sin(\hat{\vartheta}_{h}))\right)},\label{eqn:beampattern}
\end{align}
where $\omega$ spans the scanning angle range $[-\pi/2, \pi/2]$, and $d=\lambda/2$, without loss of generality. Here, $N = N_x$ is assumed for simplicity. Eq. \eqref{eqn:beampattern} demonstrates the energy beam gain for any IoT device located at $\omega$ when the beam is directed at $\hat{\vartheta}_{h}$. Thus, the beam widths on either side of the beam pattern directed towards $\hat{\vartheta}_{h}$ under a given threshold $\gamma$ are represented by
\begin{align}
    B_{\gamma, L}(\hat{\vartheta}_{h}) & = \text{width}\left\{\omega \left| \left|F_{\hat{\vartheta}_{h}}(\omega)\right|^2 \geq \gamma, \omega \leq \hat{\vartheta}_{h} \right.\right\},\nonumber\\
    B_{\gamma, R}(\hat{\vartheta}_{h}) & = \text{width}\left\{\omega \left| \left|F_{\hat{\vartheta}_{h}}(\omega)\right|^2 \geq \gamma, \omega > \hat{\vartheta}_{h} \right.\right\}.\label{eqn:beamwidth}
\end{align}

Consequently, the beam rotation scheme comprises two primary tasks: determining the optimal threshold $\gamma$ for uniform beam stitching and selecting beam directions $\hat{\vartheta}_{h}$ based on Eq. \eqref{eqn:beampattern}-\eqref{eqn:beamwidth}, aiming for complete spatial coverage. The threshold search process is detailed in Algorithm \ref{alg:2}.

\begin{figure}[!t]
	\centering
	\subfloat[First beam rotation scenario.]{\includegraphics[width=0.65\linewidth]{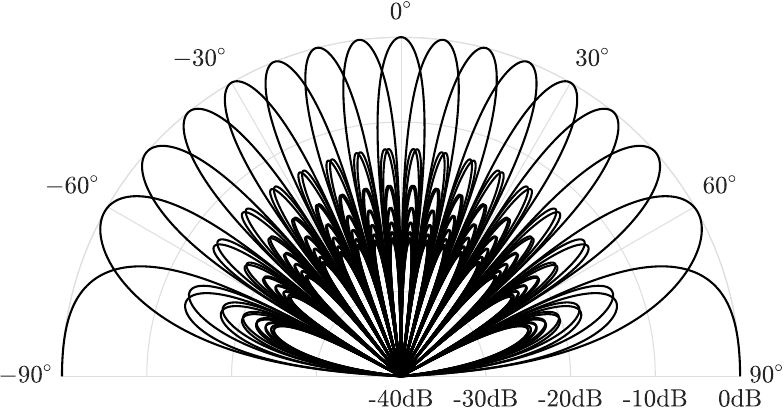}
	\label{fig:beamrotation1}}
	\setlength{\abovecaptionskip}{0pt}
	\setlength{\belowcaptionskip}{0pt}

	\subfloat[Second beam rotation scenario.]{\includegraphics[width=0.65\linewidth]{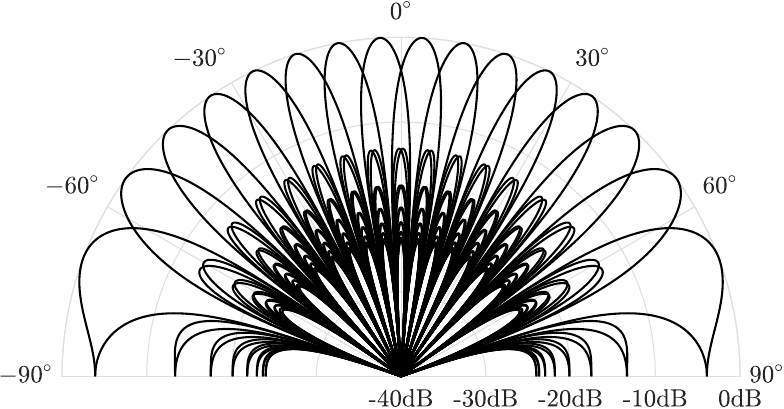}
	\label{fig:beamrotation2}}
	\setlength{\abovecaptionskip}{0pt}
	\setlength{\belowcaptionskip}{0pt}
	\caption{Illustration of two beam rotation coverage scenarios, where $N_x = 16$ and $N_B = 16$.}
	\label{fig:beamrotation}
\end{figure}

Each determined threshold $\gamma$ within the interval $[\Delta_s, \Delta_e]$ iteratively stitches beams, employing the residual left after the final beam near $0^\circ$ as a metric to determine the appropriate number of thresholds. This process, termed as the rough search, is followed by a detailed search within each feasible sub-interval $[\Delta_{s,i}, \Delta_{e,i}]$, refining the thresholds until convergence is achieved or maximum iterations are reached. The optimally obtained thresholds $\gamma_{\text{peak},i},\forall i\in N_p$, are then used for iterative beam stitching, as outlined in steps 6-9 in Algorithm \ref{alg:2}, culminating in the final beam rotation strategy. Fig. \ref{fig:beamrotation} demonstrates two feasible beam rotation schemes, where $N_B$ is the total number of beams in a single period. Note that the beam is directed at a specific direction at a time and covers the entire space over a period by beam rotation.

\subsection{Optimal Order for Beam Rotation}\label{sec:optOrder}
\begin{figure}
	\setlength{\abovecaptionskip}{0pt}
    \setlength{\belowcaptionskip}{0pt} 
	\centering
	\includegraphics[width=0.95\linewidth]{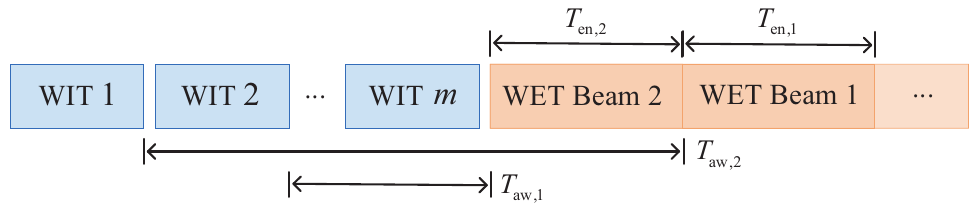}
	\caption{The definition of waiting costs.}
	\label{fig:waitingcost}
\end{figure}

Upon finalizing the beam rotation strategy, we shift focus to minimizing the waiting cost for charging, $T_\text{aw}$, defined as the average interval between the end of uplink WIT from an IoT device and the start of receiving downlink WET, as shown in Fig. \ref{fig:waitingcost} and Eq. \eqref{eqn:totalTen}-\eqref{eqn:Ten}. $T_{\text{en}, j}, \forall j\in N_B$ denotes the charging time in the $j^\text{th}$ beam. Note that we only optimize the order of beam rotation, which will not affect $T_{\text{en},j},\forall j \in N_B$, and $T_{\text{en},j}$ can be expressed as\footnote{Note that we ignore the impact of energy harvester for the sake of clarity.} 
\begin{align}
	&T_{\text{en},j}=\max \left(\frac{Q_{i}}{\sum_j PN^2M\left|F_{\hat{\vartheta}_{h,j}}(\omega_{i,j})\right|^2\varrho_{H2R}\varrho_{R2U,\{i,j\}}}\right),\nonumber\\
	&\qquad \qquad \qquad \qquad \qquad \qquad \qquad\forall i\in I_j, \forall j\in N_B,\label{eqn:Ten_j}
\end{align} 
where $Q_i$ represents the energy consumption of $i^\text{th}$ device and $I_j,\forall j \in N_B$ represents the number of IoT devices in $j^\text{th}$ beam. $\hat{\vartheta}_{h,j},\forall j \in N_B$ denotes the $j^\text{th}$ beam direction obtained from Algorithm \ref{alg:2}. $\omega_{i,j}$ represents the direction of $i^\text{th}$ device within the $j^\text{th}$ beam. $\varrho_{R2U,\{i,j\}}$ denotes the pathloss from the $i^\text{th}$ device within $j^\text{th}$ beam to the RISS. We assume that the energy consumption is the same across all devices, i.e., $Q_i=Q, \forall i\in I_j,j\in N_B$. Thus, $T_{\text{en}, j}$ can be considered associated with the worst IoT device (i.e., suffering more pathloss or less beam gain) in the corresponding beam, i.e., $T_{\text{en},j} \propto \max\left(\frac{1}{F_{\text{total}}(\omega_{i,j})\varrho_{R2U,\{i,j\}}}\right),\forall i\in I_j$\cite{CSIfreeantennaRotation}, where $F_{\text{total}}(\omega_{i,j})=\sum_j\left|F_{\hat{\vartheta}_{h,j}}(\omega_{i,j})\right|^2$.

Typically, waiting costs are considered on a per-beam basis without accounting for the number of IoT devices served by each beam. This is because all devices within a beam start and finish charging simultaneously. However, our beam rotation strategy also incorporates the impact of the IoT device count, aiming to minimize the collective waiting or average cost across all devices. This approach optimizes the charging sequence by prioritizing users with shorter required charging durations and addressing larger clusters of users to promptly meet massive charging demands. Thus, the total cost for a period, with beam rotation order $B=[b_1,b_2,\cdots,b_{N_B}]$, can be expressed as 
\begin{align}
    T_\text{aw,total} &= \tau_{b_1} + \tau_{b_2} + I_{b_2}T_{\text{en},b_1} + \cdots + \tau_{b_{N_B}}\nonumber\\
	&\qquad\qquad\qquad\qquad+I_{b_{N_B}}\sum_{k \in B, k \neq b_{N_B}} T_{\text{en},k} \nonumber \\
    &= \sum_{k \in B} \tau_k + \sum_{m=2}^{N_B} \left( I_{b_m} \sum_{k=b_1}^{b_{m-1}} T_{\text{en},k} \right),\label{eqn:totalTen}\\
    T_\text{aw} &= \frac{T_\text{aw,total}}{\sum_{j=1}^{N_B} I_j}.\label{eqn:Ten}
\end{align}
where $\tau_k, \forall k \in B$ denotes the initial delay for IoT devices in the $k^\text{th}$ beam at the onset of downlink WET. Note that we assume access to the direction information of the IoT devices in WET, which is obtained through a period of DOA estimation in WIT. This allows for the effective determination of the number of IoT devices across various beam service intervals. In other words, we have the access to $I_j,\forall j\in N_B$.

Observe from Eq. \eqref{eqn:Ten} that $\sum_{k \in B} \tau_k$ remains constant irrespective of the rotation order, directing our optimization efforts towards minimizing the latter term of Eq. \eqref{eqn:totalTen}. 

\begin{Lemma}\label{lemma:lemma2}
    The necessary and sufficient conditions for minimizing $\sum_{i=2}^{N_B} \left( I_i \sum_{j=1}^{i-1} T_{\text{en},j} \right)$ are
    \begin{align}
        \frac{I_1}{T_{\text{en},1}} > \frac{I_2}{T_{\text{en}, 2}} > \cdots > \frac{I_{N_B}}{T_{\text{en},N_B}},
    \end{align}
    implying that beams should be charged in descending order of the ratio between the number of IoT devices ($I_j,\forall j \in N_B$) in a beam and the required charging time ($T_{\text{en},j},\forall j \in N_B$) within that beam.
\end{Lemma}

\begin{IEEEproof}
    Please refer to Appendix \ref{app:B} for detailed proof.
\end{IEEEproof}

\begin{Rem}\label{remark:3}
    Lemma \ref{lemma:lemma2} elucidates the necessary and sufficient conditions to minimize waiting times, facilitating the use of information gathered by the RISS in the uplink to sequence the beam rotation order, thereby minimizing the waiting cost for WET. This strategy is universally applicable regardless of the distribution or quantity of IoT devices, enhancing its relevance across various deployment scenarios. In cases where IoT devices have $I_i = I_j, \forall i, j \in N_B$ or $T_{\text{en},i} = T_{\text{en},j}, \forall i, j \in N_B$, the conditions may simplify to charging in ascending order of $T_{\text{en},j}, \forall j \in N_B$ or descending order of $I_j, \forall j \in N_B$, respectively.
\end{Rem}

\begin{figure}
	\setlength{\abovecaptionskip}{0pt}
    \setlength{\belowcaptionskip}{0pt} 
	\centering
	\includegraphics[width=0.85\linewidth]{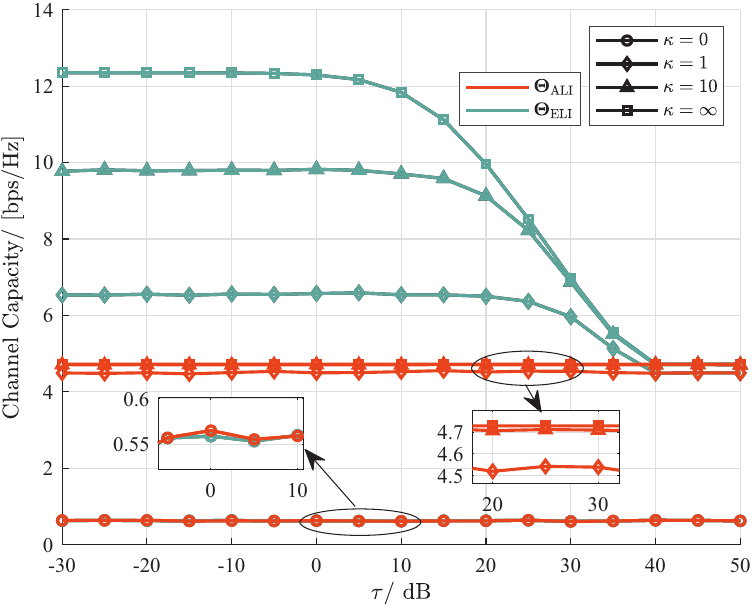}
	\caption{Performance comparison of RISS-assisted system in a Rician channel environment. The target and interference signals incident from 12.1$^\circ$ and 21.3$^\circ$ / -12.5$^\circ$, respectively.}
	\label{fig:expb1}
\end{figure}

\section{Numerical Results}\label{sec:numericalresult}
This section presents numerical results demonstrating the efficacy of our proposed RISS-assisted system for WPCNs. The distance between the HAP and the RISS $d_{R2H}$ is set at 20 meters. Signal attenuation at a reference distance of 1 meter is 30 dB, with a path loss exponent of 2.2 applied to both HAP-to-RISS and RISS-to-IoT device channels. $d_{R2I}$ ($d_{R2T}$) denotes the distance between the RISS and the interference source (target source). For simplicity, we assume uniform suppression threshold $\tau_k = \tau, \forall k \in K, k \neq d$. The system configuration includes $M = 4$ antennas at the HAP, $N = 256$ elements at the RISS, distances to the target and interference sources set to 10 meters each, noise power at $\sigma_0 = -80$ dBm, and transmit powers for uplink WIT and downlink WET set at 15.5 mW and 4 W \cite{IoTdevice}, respectively.

We first evaluate the system in a complex Rician channel environment described by:
\begin{align}
	&\mathbf{G}_\text{Rician} = \sqrt{\frac{\kappa_G}{1+\kappa_G}}\mathbf{G}+\sqrt{\frac{1}{1+\kappa_G}}\hat{\mathbf{G}},\nonumber\\
	&\mathbf{h}_{\text{Rician}, k} = \sqrt{\frac{\kappa_{h,k}}{1+\kappa_{h,k}}}\mathbf{h}_k+\sqrt{\frac{1}{1+\kappa_{h,k}}}\hat{\mathbf{h}}_k,\label{eqn:ricianchannel}
\end{align}
where $\kappa_G$ and $\kappa_{h,k}$ represent the Rician factors for the HAP-to-RISS and RISS-to-IoT channels, respectively. $\hat{\mathbf{G}}$ and $\hat{\mathbf{h}}_k$ are assumed to be circularly symmetric complex white Gaussian noise with $[\hat{\mathbf{G}}]_{i,j} \sim \mathcal{CN}(0, 1)$ for all $i \in N, j \in M$, and $[\hat{\mathbf{h}}_k]_{i} \sim \mathcal{CN}(0, 1)$ for all $i \in N$. 

As shown in Fig. \ref{fig:expb1}, the interference elimination capability of $\boldsymbol{\Theta}_{\text{ELI}}$ surpasses that of $\boldsymbol{\Theta}_{\text{ALI}}$ under conditions where $\kappa_h = \kappa_G \neq 0$ and a feasible threshold $\tau$, resulting in a higher channel capacity. With relaxed thresholds (i.e., $\tau \geq 40$ dB), the constraint specified in Eq. \eqref{eqn:eqnP1p1consC} becomes ineffective, and $\boldsymbol{\Theta}_{\text{ELI}}$ converges towards $\boldsymbol{\Theta}_{\text{ALI}}$. Additionally, the proposed RISS scheme benefits significantly from a strong LoS path, which is intuitively evident given that the RISS-based system inherently senses the LoS channel. The Rician factors $\kappa_G$ and $\kappa_{h,k}$ serve as indicators of the extent to which the active elements capture the channel characteristics. As $\kappa_G$ and $\kappa_{h,k}$ approach infinity, the Rician model converges to the deterministic channel models described in Eq. \eqref{eqn:channelG} and Eq. \eqref{eqn:channelh}, such that, in the absence of sensing errors, the channel is perfectly known. Conversely, when $\kappa_G$ and $\kappa_{h,k}$ approach zero, it indicates that the RISS has not acquired any channel information.

\begin{figure}
	\setlength{\abovecaptionskip}{0pt}
    \setlength{\belowcaptionskip}{0pt} 
	\centering
	\includegraphics[width=0.85\linewidth]{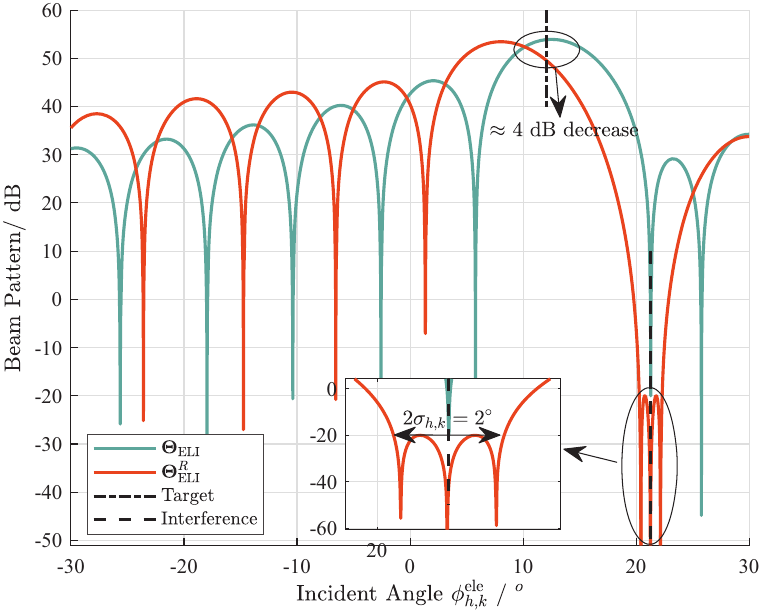}
	\caption{Robust design performance comparison under various threshold settings, where $\tau_k = -20$ dB, $N = 256$, $M = 4$, and $\delta_{h,k} = 1^\circ$. The target and interference signals incident from 12.1$^\circ$ and 21.3$^\circ$, respectively.}
	\label{fig:exp1}
\end{figure}

\begin{figure}
	\setlength{\abovecaptionskip}{0pt}
    \setlength{\belowcaptionskip}{0pt} 
	\centering
	\includegraphics[width=0.85\linewidth]{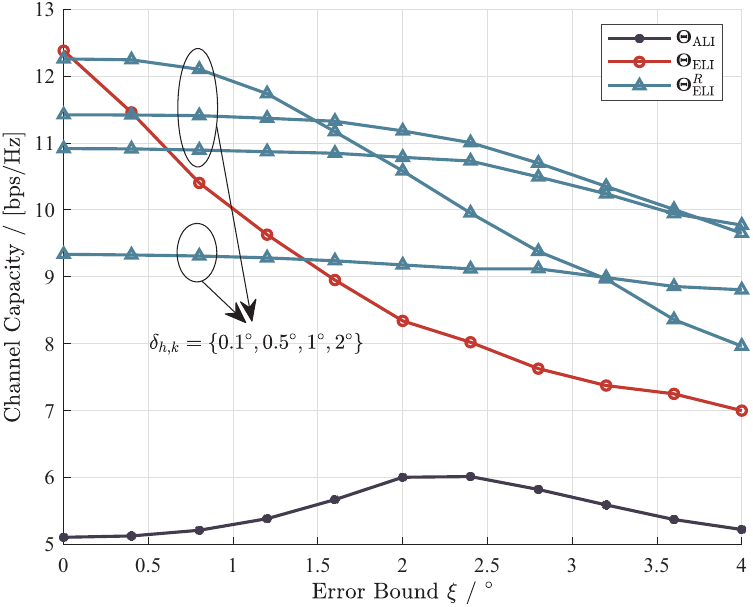}
	\caption{Monte Carlo simulations of channel capacity under different interference suppression strategies. The angle estimation errors are uniformly distributed within $[-\xi, \xi]$, where $d_{R2T} = d_{R2I} = 10$ meters.}
	\label{fig:exp1-2}
\end{figure}

Furthermore, the impact of the robust design introduced in Section \ref{sec:robustdesign} is depicted in Fig. \ref{fig:exp1}. The results indicate that $\boldsymbol{\Theta}^R_\text{ALI}$, compared to $\boldsymbol{\Theta}_\text{ALI}$, broadens the nulls of interference suppression effectively within the angle range $[\hat{\phi}^\text{ele}_{h,k} - \delta_{h,k}, \hat{\phi}^\text{ele}_{h,k} + \delta_{h,k}]$, determined by $\tau_k, k \in K, k \neq d$ and the resolution of constraints. Despite a slight decrease in the gain towards the target direction, the improved interference suppression leads to a net increase in SINR. The relationship between SINR and the sensing error of signal angles is further explored in Fig. \ref{fig:exp1-2}, revealing a significant enhancement in the system's overall SINR despite the presence of estimation errors.

Fig. \ref{fig:exp1-2} illustrates that $\boldsymbol{\Theta}_\text{ELI}$ outperforms $\boldsymbol{\Theta}^R_\text{ELI}$ when the error is small (e.g., $\xi=0^\circ$), and its performance diminishes as the error increases. Furthermore, appropriately chosen $\tau_k, k\in K, k\neq d$ can yield consistent performance despite increasing $\xi$. However, excessive suppression (e.g., $\delta_{h,k}=2^\circ$) adversely impacts channel capacity in scenarios with small errors. Note that despite the error surpassing the range accounted for in the robust design, the channel capacity experiences a gradual decrease. This outcome is attributed to the robust design's ability to mitigate the sharp rise on either side of the beam pattern nulls, thereby slowing down the decrease in channel capacity. Another intriguing observation is the behavior of the channel capacity for $\boldsymbol{\Theta}_\text{ALI}$, which demonstrates an initial increase followed by a decrease with rising $\xi$, attributed to the interference signal sliding more rapidly to nulls within the side lobes compared to the target signal within the main lobe.

\begin{figure}
	\setlength{\abovecaptionskip}{0pt}
    \setlength{\belowcaptionskip}{0pt} 
	\centering
	\includegraphics[width=0.92\linewidth]{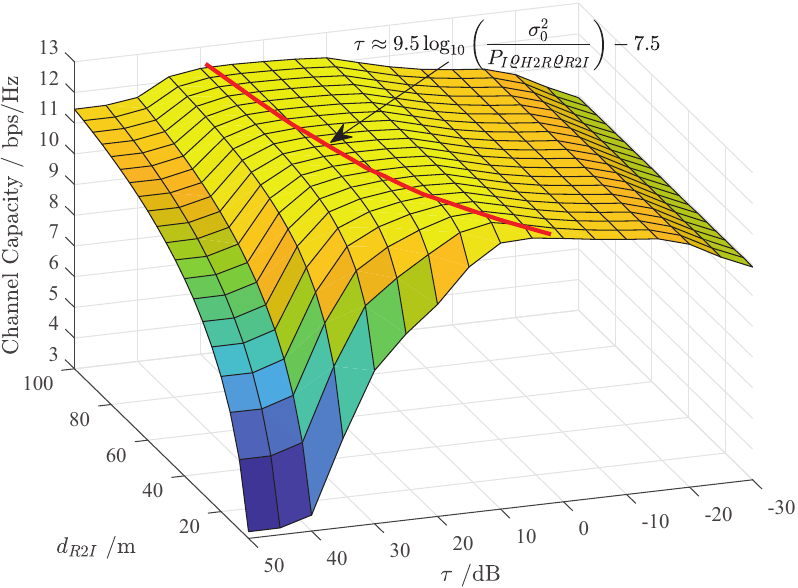}

	\caption{The channel capacity varies with distance $d_{R2I}$ and suppression threshold $\tau$, where the distance between the RISS and the target source is set to $d_{R2T}=10$ meters.}
	\label{fig:exp2}
\end{figure}

Fig. \ref{fig:exp2} illustrates the channel capacity variation concerning the distance $d_{R2I}$ and the various suppression threshold $\tau$. Note that for the interference sources closer to the RISS (e.g, $d_{R2I}=10$ meters, corresponding to higher-power interference signals), a preference leans toward employing smaller suppression thresholds $\tau$ to effectively contain the interference. However, for interference sources farther away from the RISS (e.g., $d_{R2I}=100$ meters, corresponding to lower-power interference signals), it is advisable to utilize more relaxed suppression thresholds. This recommendation arises from the fact that excessively small suppression constraints can lead to a loss in the gain of the target direction, thereby compromising the overall SINR. This becomes particularly critical when the interference signal power and noise power are within a similar magnitude, potentially resulting in detrimental outcomes due to aggressive interference suppression. We offer an approximate expression to achieve the maximum channel capacity, represented as
\begin{align}
	\tau \approx 9.5\log_{10}\left(\frac{\sigma_0^2}{P_I\varrho_{H2R}\varrho_{R2I}}\right)-7.5,
\end{align}
where $\varrho_{H2R}\varrho_{R2I}$ represents the cascaded path loss between the HAP and the interference source, correlating with the distance $d_{H2R}$ and $d_{R2I}$.

\begin{figure}
	\setlength{\abovecaptionskip}{0pt}
    \setlength{\belowcaptionskip}{0pt} 
	\centering
	\includegraphics[width=0.7\linewidth]{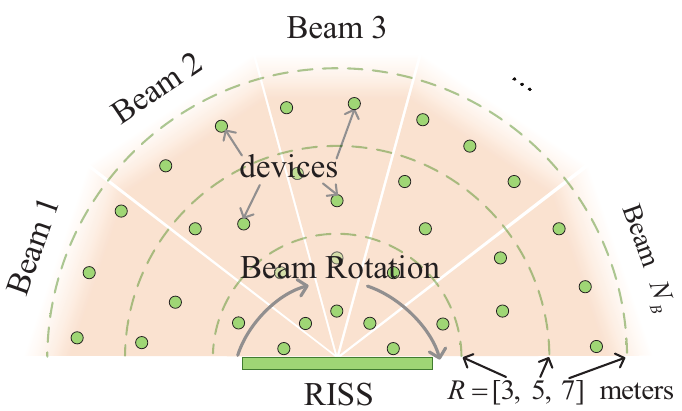}
	\caption{Example devices deployments for WET.}
	\label{fig:devicelocation}
\end{figure}


We introduce the practical non-linear energy harvesting model\cite{NonLinearEH}, which captures the dynamics of RF energy conversion efficiency at various input power levels. Specifically, the total harvested energy at $i^{\text{th}}$ IoT device can be expressed as
\begin{align}
	f(P_i) &= \frac{M_s}{X(1+\exp(-a(P_i-b)))}-Y, \quad\left[\text{Watt}\right]\nonumber\\
	X &= \frac{\exp(ab)}{1+\exp(ab)}, \quad Y= \frac{M_s}{\exp(ab)},
\end{align}
where $f(\cdot)$ denotes the non-linear relationship between the input RF power and the harvested energy, and $P_i$ is the input RF power of the $i^{\text{th}}$ IoT device. The parameters $a=132.8$, $b=0.01181$, and $M_s=0.02337$ \cite{6712143} represent the joint impact of the resistances, capacitances, and circuit sensitivity on the rectifying process.

\begin{figure}
	\setlength{\abovecaptionskip}{0pt}
    \setlength{\belowcaptionskip}{0pt} 
	\centering
	\includegraphics[width=0.85\linewidth]{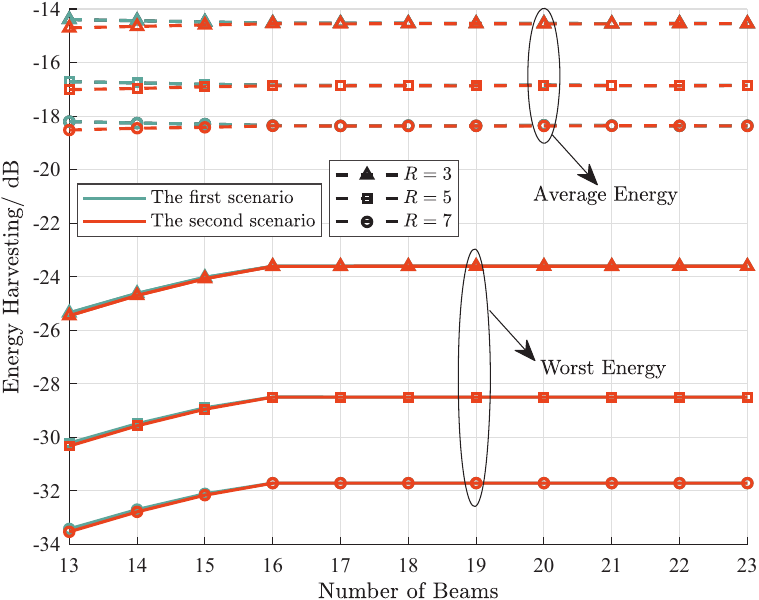}

	\caption{The Monte Carlo simulations of average and worst energy performance of two scenarios, where $S=50$ IoT devices are uniformly distributed in the semi-circular area of $R=[3, 5, 7]$ meters.}
	\label{fig:WETexp1}
\end{figure}

Fig. \ref{fig:WETexp1} illustrates the received energy performance of two scenarios introduced in Fig. \ref{fig:beamrotation}, and the charging time of each beam is $T_{\text{en}, j}=1/N_B,\forall j \in N_B$ for clarity. Note that both scenarios exhibit similar average performance and worst energy performance\footnote{The worst energy performance is defined as the minimal harvested energy in the $S$ IoT devices, which can better characterizes the fairness of energy performance.} when $S=50$ IoT devices are uniformly distributed within the semi-circular area of $R=[3, 5, 7]$ meters, as shown in Fig. \ref{fig:devicelocation}. For the sake of convenience, we will adopt the first scheme for the subsequent experiments. Additionally, it is observed that the worst energy performance of two schemes gradually flat, suggesting the necessity for $N_B\ge16$ beams to cover the entire space under this configurations.

To further highlight the benefits from sensing for massive WET, we assume the device number of each beam is uniformly distributed in range $[1, \mathcal{N}]$, and the locations of devices in each beam are uniformly distributed. Thus, the number of devices in each beam forms a non-uniform distribution of devices clusters. And the operation current and supply voltage of ultra-low power consumption IoT devices are shown in Table. \ref{table:IoTpara}, where sensing including humidity and temperature measurement\cite{IoTdevice}.

\begin{table}[t]
    \centering
	\caption{The parameters of ultra-low power consumption IoT devices}
    \begin{tabular}{c|c|c|c}
    \hline
    \textbf{}&\text{Current} &\text{Duration}&\text{Supply Voltage}\\
	\hline
	\text{Standby}&1$\mu$A&-&\multirow{3}*{3.0 V}\\\cline{ 1 - 3 }
	\text{I2C Comm.\& Sensing}&10 mA& 30ms&\\\cline{ 1 - 3 }
	\text{Data Transmission}&15 mA & 10ms&\\\cline{ 1 - 3 }
	\hline
	\text{Total Consumption}&\multicolumn{3}{c}{$Q=1.35$ mJ}\\
	\hline
	\end{tabular} \label{table:IoTpara}
\end{table}

\begin{figure}
	\setlength{\abovecaptionskip}{0pt}
    \setlength{\belowcaptionskip}{0pt} 
	\centering
	\includegraphics[width=0.85\linewidth]{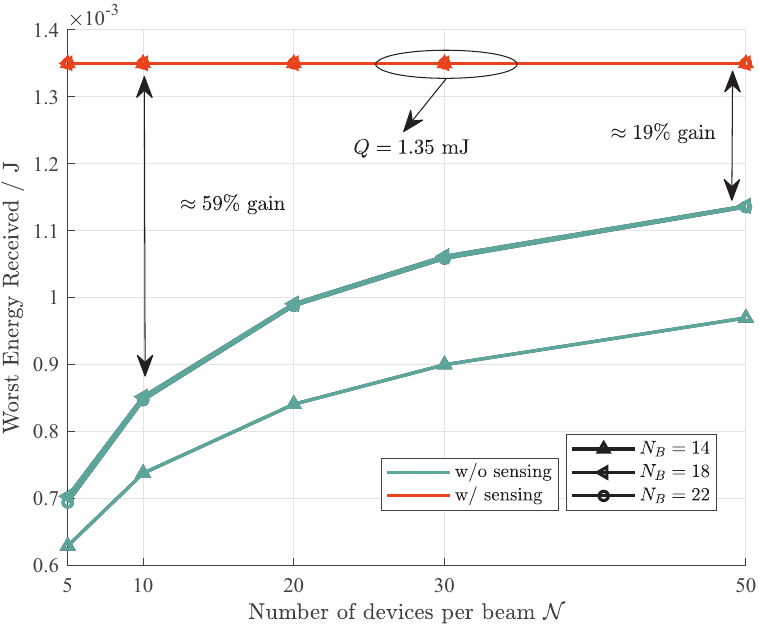}
	\caption{The worst energy performance of both with and without sensing-aided massive WET. }
	\label{fig:WETexp2}
\end{figure}

With the aid of sensing information, we can obtain $T_{\text{en},j}$ via Eq. \eqref{eqn:Ten_j}. And we set $T^{w/o}_{\text{en},j}=\sum_{j\in N_B}T_{\text{en},j}/N_B$ as the counterpart without sensing information. Fig. \ref{fig:WETexp2} demonstrates the worst energy performance with and without sensing-aided massive WET. The sensing-aided massive WET can supplement IoT device consumption with $Q=1.35$ mJ, showing an improvement of 59\% and 19\% compared to its counterpart for $\mathcal{N} = 10$ and $\mathcal{N} = 50$, respectively. These results highlight the benefits of incorporating sensing information. Moreover, the gain decreases as the number of devices per beam increases. This phenomenon occurs because, as the number of devices increases, the worst-case scenario across all beams tends to become more similar, i.e., $\max\left(\frac{1}{F_{\text{total}}(\omega_{i,j})\varrho_{R2U,\{i,j\}}}\right)=\max\left(\frac{1}{F_{\text{total}}(\omega_{l,k})\varrho_{R2U,\{l,k\}}}\right),\forall i\in I_j, j\in N_B, l\in I_k, k\in N_B$, resulting $T_{\text{en},j}=T_{\text{en},k}$. Additionally, the cases of $N_B=18$ and $N_B=22$ exhibit similar performance, both outperforming the case of $N_B=14$, presenting the same results as shown in Fig. \ref{fig:WETexp1}.

As the final simulation, we demonstrate the enhanced performance in reducing waiting costs through sensing. Note that we have access to $T_{\text{en},j}$ and $I_j,\forall j\in N_B$ via sensing, and due to $T_{\text{en}}\gg \sum _{k\in B}\tau_k$, we ignore the first component of Eq. \eqref{eqn:totalTen}. Consequently, the optimal order for the proposed massive WET is obtained from Lemma \ref{lemma:lemma2}, while the counterpart order is sequential from $1^\text{th}$ to $N_B^\text{th}$.

\begin{figure}
	\setlength{\abovecaptionskip}{0pt}
    \setlength{\belowcaptionskip}{0pt} 
	\centering
	\includegraphics[width=0.85\linewidth]{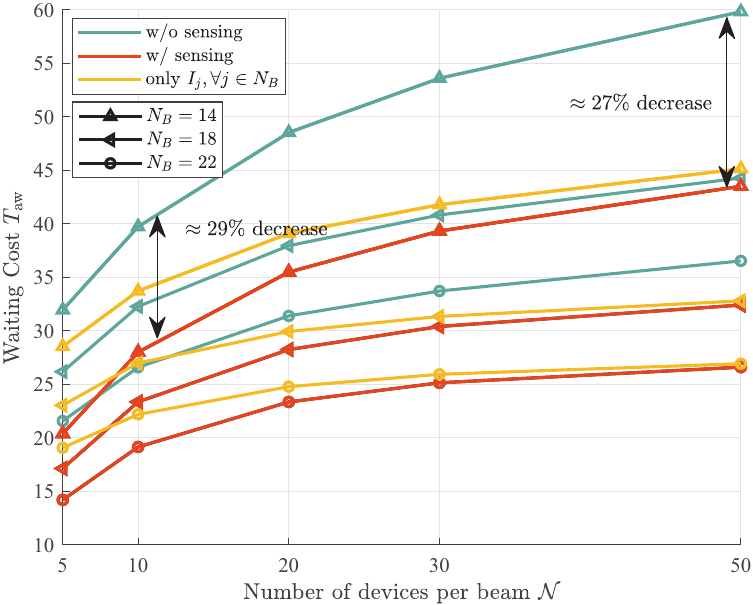}
	\caption{The waiting cost of both with and without sensing-aided massive WET. }
	\label{fig:WETexp3}
\end{figure}

Fig. \ref{fig:WETexp3} shows a significant decrease in waiting costs, by 29\% and 27\% when $\mathcal{N}=10$ and $\mathcal{N}=50$, respectively, highlighting the benefits of sensing. We also plot an additional case where only $I_j,\forall j\in N_B$ is available, with the order following the descending values of $I_j$. As $\mathcal{N}$ increases, the performance between the optimal and the order based solely on $I_j$ becomes similar. This is due to the fact that $T_{\text{en},j}=T_{\text{en},k}$ as the number of devices increases, simplifying the optimal order to the descending order of $I_j,\forall j\in N_B$, as mentioned in Remark \ref{remark:3}.

\section{Conclusion}\label{sec:conclusion}

This paper proposed a novel RISS-assisted WPCN system. By utilizing angle estimation and signal identity determination from the active elements of RISS, we designed an interference suppression WIT algorithm for RISS passive beamforming in the uplink and a massive WET algorithm in the downlink. Motivated by the effectiveness of interference suppression, we also developed a robust algorithm for scenarios with imperfect angle estimation. Our findings demonstrate that the WPCN system significantly benefits from both single sensing (for WIT) and multi-sensing (for WET), leading to substantial improvements in SINR in the uplink and providing efficient WET performance. Additionally, we determined the optimal threshold for interference suppression to ensure that WIT performance is not compromised by excessive suppression. The charging time and beam rotation order for massive WET are optimized to enhance the worst-case performance and minimize waiting costs.

{\appendices
\section{Proof of Lemma \ref{lemma:lemma1}}\label{app:A}
To prove this conclusion, we first rewrite the interference signal as
\begin{align}
	&\left|\mathbf{v}^T\mathbf{G}^T\boldsymbol{\Theta}\mathbf{h}_k\right|^2\nonumber\\
	&=\left|\mathbf{v}^T\boldsymbol{\beta}(\varpi_{G})\text{diag}\{\boldsymbol{\Theta}\}^T\left(\boldsymbol{\alpha}(\vartheta_{G}, \varphi_{G})\circ\boldsymbol{\alpha}(\vartheta_{h,k}, \varphi_{h,k})\right)\right|^2\nonumber\\
	&\overset{(a)}{=}\left|\mathbf{v}^T\boldsymbol{\beta}(\varpi_{G})\text{diag}\{\boldsymbol{\Theta}\}^T\left(\boldsymbol{\alpha}_x(\varphi_G)\otimes \boldsymbol{\alpha}_y(\vartheta_G)\right)\right.\nonumber\\
		&\qquad \qquad \qquad \qquad \qquad \qquad \quad \left.\circ\left(\boldsymbol{\alpha}_x(\varphi_{h,k})\otimes \boldsymbol{\alpha}_y(\vartheta_{h,k})\right)\right|^2\nonumber\\
	&\overset{(b)}{=}PM\left|\left(\boldsymbol{\alpha}_x(\varphi_{G}+\varphi_{h,k})^H\boldsymbol{\alpha}_x(\varphi_{G}+\varphi_{h,d})\right)\right.\nonumber\\
	&\left.\qquad \qquad \qquad \qquad \otimes\left(\boldsymbol{\alpha}_y(\vartheta_{G}+\vartheta_{h,k})^H\boldsymbol{\alpha}_y(\vartheta_{G}+\vartheta_{h,d})\right)\right|^2\nonumber\\
	&\overset{(c)}{=}PM\left|\mathbf{1}^T_N\left(\boldsymbol{\alpha}_x(\varphi_{h,d}-\varphi_{h,k})\otimes\boldsymbol{\alpha}_y(\vartheta_{h,d}-\vartheta_{h,k})\right)\right|^2,\label{eqn:interferencesignal}
\end{align}
where $(a)$ comes from $\boldsymbol{\alpha}_x(\varphi)\otimes\boldsymbol{\alpha}_y(\vartheta)=\boldsymbol{\alpha}(\vartheta, \varphi)$. $(b)$ comes from Eq. \eqref{eqn:alignv} and Eq. \eqref{eqn:aligntheta}, $(\mathbf{A}\otimes\mathbf{B})^H=\mathbf{A}^H\otimes\mathbf{B}^H$, $(\mathbf{A}\otimes\mathbf{B})\circ(\mathbf{C}\otimes\mathbf{D})=(\mathbf{A}\circ\mathbf{C})\otimes(\mathbf{B}\circ\mathbf{D})$ and $(\mathbf{A}\otimes\mathbf{B})(\mathbf{C}\otimes\mathbf{D})=(\mathbf{A}\mathbf{C})\otimes(\mathbf{B}\mathbf{D})$. To eliminate interference signals, we have
\begin{align}
	\mathbf{1}^T_N\left(\boldsymbol{\alpha}_x(\varphi_{h,d}-\varphi_{h,k})\otimes\boldsymbol{\alpha}_y(\vartheta_{h,d}-\vartheta_{h,k})\right)=0,
\end{align}
which can be rewritten as
\begin{align}
	&\mathbf{1}^T_N\left(\boldsymbol{\alpha}_x(\varphi_{h,d}-\varphi_{h,k})\otimes\boldsymbol{\alpha}_y(\vartheta_{h,d}-\vartheta_{h,k})\right)\nonumber\\
	&\overset{(a)}{=}\left(\mathbf{1}_{N_x}^T\boldsymbol{\alpha}_x(\varphi_{h,d}-\varphi_{h,k})\right)\left(\mathbf{1}_{N_y}^T\boldsymbol{\alpha}_y(\vartheta_{h,d}-\vartheta_{h,k})\right)\nonumber\\
	&=0,\label{eqn:intereliminate}
\end{align}
where $(a)$ comes from $\mathbf{1}^T_{N}=\mathbf{1}^T_{N_x}\otimes\mathbf{1}^T_{N_y}$ and $(\mathbf{A}\otimes\mathbf{B})(\mathbf{C}\otimes\mathbf{D})=(\mathbf{A}\mathbf{C})\otimes(\mathbf{B}\mathbf{D})$. Since
\begin{align}
	&\mathbf{1}_{N_x}^T\boldsymbol{\alpha}_x(\varphi_{h,d}-\varphi_{h,k})=\frac{1-e^{\mathbbm{i}\pi N_x\left(\cos\left(\phi^{\text{azi}}_{h,d}\right)-\cos\left(\phi^{\text{azi}}_{h,k}\right)\right)}}{1-e^{\mathbbm{i}\pi \left(\cos\left(\phi^{\text{azi}}_{h,d}\right)-\cos\left(\phi^{\text{azi}}_{h,k}\right)\right)}},
\end{align}
an essential requirement for Eq. \eqref{eqn:intereliminate} to hold is
\begin{align}
	\left|\cos(\phi^{\text{azi}}_{h,d})-\cos(\phi^{\text{azi}}_{h,k})\right|=\frac{2n}{N_x},n\in\mathbb{N}^+.
\end{align}
Similarly, the another essential requirement for Eq. \eqref{eqn:intereliminate} to hold is
\begin{align}
	\left|\sin(\phi^{\text{azi}}_{h,d})\sin(\phi^{\text{ele}}_{h,d})-\sin(\phi^{\text{azi}}_{h,k})\sin(\phi^{\text{ele}}_{h,k})\right|=\frac{2n}{N_y},n\in\mathbb{N}^+.
\end{align}
Up to now, we derive the outcome presented in Lemma \ref{lemma:lemma1}.
\section{Proof of Lemma \ref{lemma:lemma2}} \label{app:B}
	We first prove the sufficient conditions. While the order $B=[b_1,\cdots,b_{N_B}]$ satisfies $\frac{I_{b_1}}{T_{b_1}}>\frac{I_{b_2}}{T_{b_2}}>\cdots>\frac{I_{b_{N_B}}}{T_{b_{N_B}}}$, we randomly exchange the $m^\text{th}$ and $n^\text{th}$ order in $B$, and $m<n$ for the clarity. While $n-m=1$, the latter component of Eq. \eqref{eqn:totalTen} can be rewritten as
	\begin{align}
		&C^{m, m+1}_{\text{ex}}=I_{b_2}T_{\text{en},b_1}+\cdots+I_{b_{N_B}}\sum_{k\in N_B,k\neq b_{N_B}}T_{\text{en},k}\nonumber\\
		&=\cdots+I_{b_{m+1}}\left(T_{\text{en},b_1}+\cdots+T_{\text{en}, b_{m-1}}\right)\nonumber\\
		&\qquad+I_{b_m}\left(T_1+\cdots+T_{\text{en}, b_{m-1}}+T_{\text{en}, b_{m+1}}\right)+\cdots.\label{eqn:ex_adjacent}
	\end{align}
	Thus, the difference between Eq. \eqref{eqn:ex_adjacent} and original $C=\sum_{i=2}^{N_B}\left(I_i\sum_{j=1}^{i-1}T_j\right)$ can be expressed as
	\begin{align}
		C-C^{m, m+1}_{\text{ex}} &= I_{b_{m+1}}T_{\text{en},b_m}-I_{b_{m}}T_{\text{en},b_{m+1}}\nonumber\\
		&=I_{b_{n}}T_{\text{en},b_m}-I_{b_{m}}T_{\text{en},b_{n}}.
	\end{align}
	According to the relationship $n>m$ and $\frac{I_{b_{m}}}{T_{\text{en},b_m}}>\frac{I_{b_{n}}}{T_{\text{en},b_n}}$, we can derive
	\begin{align}
		C-C^{m, n}_{\text{ex}}<0,
	\end{align}
	which demonstrates the increase in the total waiting time, $T_{\text{aw, total}}$.
	
	Similarly, while $n-m>1$, the difference between $C$ and $C_\text{ex}$ can be expressed as
	\begin{align}
		&C-C^{m, n}_{\text{ex}}  \nonumber\\
		=&\sum_{j=m}^{n-1}\left(I_{b_n}T_{\text{en}, b_j}-I_{b_j}T_{\text{en},b_n}\right)\nonumber\\
		&\qquad\quad\qquad\qquad+\sum_{i=m+1}^{n-1}\left(I_{b_i}T_{\text{en},b_m}-I_{b_m}T_{\text{en},b_i}\right),
	\end{align}
	since $j<n,\forall j\in[m,n-1]$ and $i>m, \forall i\in[m+1,n-1]$ holds, we further have
	\begin{align}
		\frac{I_{b_j}}{T_{\text{en},b_j}}>\frac{I_{b_n}}{T_{\text{en},b_n}}, \forall j\in[m,n-1],\nonumber\\
		\frac{I_{b_m}}{T_{\text{en},b_m}}>\frac{I_{b_{i}}}{T_{\text{en},b_{i}}}, \forall i\in[m+1,n-1],
	\end{align}
	which illustrate the fact that $C-C^{m,n}_\text{ex}<0$, leading to an increase in $T_{\text{aw, total}}$. Hence, the sufficient conditions have been established.

	As for the necessary conditions, assuming the rotation of beams follows the order $B=[b_1,\cdots,b_{N_p}]$ to achieve the minimum total waiting time. Under this premise, interchanging the sequence of any adjacent two beams (e.g., $b_k$ and $b_{k+1}$) will inevitably increase the overall duration, thereby obtaining
	\begin{align}
		&C-C_{\text{ex}}^{b_k,b_{k+1}}<0\nonumber\\
		&\to I_{b_k}\sum_{i=1}^{k-1}T_{\text{en},b_i}+I_{b_{k+1}}\sum_{i=1}^{k}T_{\text{en},b_i}-I_{b_{k+1}}\left(\sum_{i=1}^{k-1}T_{\text{en},b_i}\right)\nonumber\\
		&\qquad\qquad\qquad\qquad -I_{\text{en},b_{k}}\left(T_{\text{en},b_{k+1}}+\sum_{i=1}^{k-1}T_{\text{en},b_i}\right)<0\nonumber\\
		&\to I_{b_{k+1}}T_{\text{en},b_{k}}-I_{b_{k}}T_{\text{en},b_{k+1}}< 0,
	\end{align}
	which can be derived into
	\begin{align}
		\frac{I_{b_{k}}}{T_{\text{en},b_{k}}}>\frac{I_{b_{k+1}}}{T_{\text{en},b_{k+1}}}.
	\end{align}
	Similarly, by examining all pairs of $(b_k, b_{k+1}),\forall k\in N_B$, we can establish the necessary conditions.
}

\bibliography{Reference}

\end{document}